\documentclass[twocolumn, twocolappendix, resetfootnote]{aastex701}
\usepackage{gensymb}
\RequirePackage{natbib}
\hypersetup{breaklinks=true, colorlinks=true,
citecolor=darkred, urlcolor=deepmagenta, linkcolor=darkblue,
bookmarks=true, bookmarksopenlevel=2}
\usepackage{multirow}
\usepackage{ulem}

\definecolor{darkblue}{rgb}{0.0, 0.0, 0.62}
\definecolor{deepmagenta}{rgb}{0.8, 0.0, 0.5}
\definecolor{darkred}{rgb}{0.55, 0.0, 0.0}
\definecolor{violet}{rgb}{0.6, 0.0, 1.0}

\RequirePackage{mymacro}

\defcitealias{sinha2026aa}{S26}
\shorttitle{Revealing Galactic dust beneath the Cosmic Infrared Background anisotropies}
\shortauthors{S.\ Sinha \etal}
\received{ }
\revised{ }
\accepted{ }
\submitjournal{ApJ}
\begin{document}
\title{Revealing Galactic dust beneath the cosmic infrared
background anisotropies with Wavelet Phase Harmonics}
\author[orcid=0000-0002-4804-3654, sname='Sinha']{Srijita Sinha} 
\affiliation{School for Physical Sciences, National Institute of Science Education and Research, HBNI Jatni-752050, India}
\affiliation{Homi Bhabha National Institute, Training School Complex, Anushakti Nagar, Mumbai 400094, India} 
\affiliation{Raman Research Institute, C.\ V.\ Raman Avenue, Sadashivanagar, Bengaluru 560080, India}
\email[show]{sinha.srijita@rrimail.rri.res.in}

\author[orcid=0000-0001-6088-3034,sname='Ghosh']{Tuhin Ghosh}
\affiliation{School for Physical Sciences, National Institute of Science Education and Research, HBNI Jatni-752050, India}
\affiliation{Homi Bhabha National Institute, Training School Complex, Anushakti Nagar, Mumbai 400094, India} 
\email[show]{tghosh@niser.ac.in}  

\author[orcid=0000-0003-3755-7593, sname='Allys']{Erwan Allys} 
\affiliation{Laboratoire de Physique de l'\'{E}cole normale sup\'erieure, ENS, Universit\'e PSL, CNRS, Sorbonne Universit\'e, Universit\'e Paris Cit\'e, F-75005 Paris, France}
\email{erwan.allys@phys.ens.fr}

\author[orcid=0000-0003-1097-6042, sname='Boulanger']{Fran\c{c}ois Boulanger} 
\affiliation{Laboratoire de Physique de l'\'{E}cole normale sup\'erieure, ENS, Universit\'e PSL, CNRS, Sorbonne Universit\'e, Universit\'e Paris Cit\'e, F-75005 Paris, France}
\email{francois.boulanger@phys.ens.fr}

\author[orcid=0000-0002-0713-1658, sname='Delouis']{Jean-Marc Delouis} 
\affiliation{Laboratoire d'Oc{\'e}anographie Physique et Spatiale (LOPS), Univ. Brest, CNRS, Ifremer, IRD, Brest, France}
\email{Jean.Marc.Delouis@ifremer.fr}
\begin{abstract}
Separating Galactic dust emission from cosmic infrared background (CIB) anisotropies is challenging because both the emission components have similar spectral properties. The primary objective of this work is to develop a component-separation algorithm to extract the Galactic dust emission from CIB anisotropies dominated sky regions using Wavelet Phase Harmonics statistics (WPH). We apply our algorithm to the \textit{Planck} $353\,\mathrm{GHz}$ frequency band over three sky regions spanning a range of neutral hydrogen ($\mathrm{\ion{H}{I}}$) column densities at high Galactic latitudes. In all three regions, we successfully recover the dust signal without significant leakage between the dust and CIB anisotropy components. We use the low signal-to-noise ($\mathrm{S/N}$) region to validate the component-separation algorithm and compare its performance with the standard template-fit approach that uses the linear dust-$\mathrm{\ion{H}{I}}$ correlation relation. For intermediate and high $\mathrm{S/N}$ regions, we separate the dust signal map from the contamination while preserving the statistical correlation with $\mathrm{\ion{H}{I}}$ column density map. Dust maps are used to analyze the relationship between dust and \HI emissions. We characterize the spatial variations in dust emission relative to the hydrogen column density as a function of the angular scale. 
\end{abstract}

\section{Introduction} \label{sec:intro}

The \Planck survey provides the best full-sky maps till date of the microwave and far-infrared sky with unprecedented sensitivity and angular resolution in nine frequency bands varying from $30 \GHz$ to $857 \GHz$~\citep{planck-I:2018}. At frequencies above $200 \GHz$, diffuse far-infrared thermal radiation from Galactic dust dominates the observed sky at large angular scales~\citep{Planck2015:XI}, while the integrated emission from distant star-forming galaxies --- the cosmic infrared background~\citep[CIB;][]{partridge1967apj}, contributes predominantly at smaller angular scales. An accurate characterisation of the dust emission is crucial for understanding the complex physical and chemical processes in the interstellar medium (ISM;\citealt{planck-XIX:2011,planck-XI:2013, planck-XVII:2014}) as well as the history of star formation and galaxy evolution~\citep{Draine_and_Li:2007}. Moreover, robust modelling of dust emission plays an important role in the analysis of supernova observations~\citep{Riess:1996} and in constraining the amplitude of the $B$-modes of the cosmic microwave background (CMB) radiation~\citep{Remazeilles:2016}. The CIB encodes the information about galaxy formation over cosmic time and traces the large-scale matter distribution of the Universe~\citep{Puget:1996, hauser1998apj, lagache1999aa, Lagache:2005sw, hauser2001araa, planck-XXX:2014, maniyar2018aa}. It provides a powerful probe of large-scale structure through galaxy clustering and weak gravitational lensing, and plays an important role in obtaining the CMB lensing potential~\citep{Larsen:2016, chluba2019ax}. Emission from Galactic dust and the CIB anisotropies not only act as contaminants for CMB studies, but also presents a formidable challenge in separating the two components. Although originating from distinct physical processes, they follow the same emission law with slightly different spectral parameters, making their separation difficult using only the spectral information. 

The heated dust grains in the optically thin medium mix well with the gas in the ISM, resulting in a tight correlation between the dust emission and the $21\,\cm$ integrated emission from the neutral hydrogen (\HI;~\citealt{Boulanger:1996, Boulanger:1988}). This linear dust-\HI correlation has been widely used in template-fitting approach to separate Galactic dust emission from the CIB anisotropies~\citep{planck-XXIV:2011, planck-XVII:2014, Lenz:2019, adak2024mnras, sinha2026aa}. However, the template-fitting approach is limited by uncertainties in the determination of the global offset level, spatial variations in dust emissivity, and the presence of excess Galactic emission that is uncorrelated with the \HI emission. Alternatively, model-independent techniques such as the generalised needlet internal linear combination exploit both spectral and spatial information to disentangle dust emission from the CIB anisotropies~\citep{planck-XLVIII:2016}, but they require prior knowledge of the spatial correlation of the CIB anisotropies at the power spectrum level. Several other approaches have also been developed to obtain unbiased and accurate estimates of the CIB power spectrum~\citep{planck-X:2015, Mak:2017, viero2019apj, chiang2019apj}.

Albeit these advances, unbiased separation of Galactic dust emission and the CIB anisotropies remains challenging, particularly in low column-density regions where the CIB anisotropies dominates the dust signal. This motivates the need for improved statistical techniques capable of reliably separating the two components. One such promising approach is the statistical component-separation, using the recently developed Scattering Transform (ST) statistics~\citep{bruna2013ieee, anden2014ieee}. ST statistics are low variance summary statistics that characterise the interactions between different scales efficiently and capture the non-Gaussian structures through wavelet convolutions and nonlinear operators~\citep{cheng2021ax}. They can be used to construct statistically identical synthetic fields for a given physical process using the maximum entropy generative model~\citep{bruna2018jmst, allys2020prd,cheng2024pnas}. Since their conception, the ST-based methods have been significantly developed and expanded, emerging as powerful tools for analyzing complex astrophysical and cosmological fields, with applications ranging from studies of the interstellar medium~\citep{allys2019aa, regaldo2020aa, saydjari2021apj, lei2023apj, richard2025aa} to different stages of cosmological evolution~\citep{allys2020prd, cheng2020mnras, cheng2021mnras, eickenberg2022ax, valogiannis2022prd,greig2024mnras, hothi2024aa},  for separating the dust polarization maps from instrumental noise in the \Planck~\citep{regaldo2021aa, delouis2022aa}, dust-CIB separation in the \herschel SPIRE~\citep{auclair2024aa} and \Planck  observations~\citep{sinha2026aa}.

Recently,~\citet[][hereafter~\citetalias{sinha2026aa}]{sinha2026aa} have applied the Scattering Covariance (SC) statistics based component-separation method to the \Planck 353 GHz frequency band in order to separate the dust emission from the contamination (dominated by the CIB anisotropies) at high Galactic latitudes ($\abs{b} > 45\degree$). That work is primarily focused on sky regions where the signal-to-noise ratio ($\SNR$) of the dust signal with respect to the contamination is high ($\SNR \geqslant 3$) and where there is a significant difference seen between the \Planck $353 \GHz$ map and the dust reddening map at $100\,\mum$ from~\cite{chiang2023apj}. The main goal of this work is to analyze those sky regions where the contamination dominates over the dust signal (or low $\SNR$ regions), using a similar ST-based component-separation method. We use wavelet phase harmonics (WPH) statistics, a specific subclass of the ST statistics, for our analysis. We improve the component-separation algorithm implemented in~\citetalias{sinha2026aa} to probe the spatial scale of dust emissivity variations at 353\,GHz. Any residual emission, even after accounting for variable dust emissivity up to the correlation scale, could hint at dust emission from a multiphase ISM (e.g.\ molecular hydrogen and/or ionised hydrogen).

The paper is organized as follows. In Sect.~\ref{sec:data-set}, we describe the datasets used in our study. In Sect.~\ref{sec:wph-algorithm}, we briefly present the component-separation algorithm using the WPH statistics. We present the main component-separation results for three different $\SNR$ regions at high Galactic latitude in Sect.~\ref{sec:data-results}. The output maps are used to analyze the relationship between dust and \HI in Sect.~\ref{sec:dust-emissivity}. Finally, our conclusions are presented in Sect.~\ref{sec:discussion}. In Appendix~\ref{app:wph-coeff}, we discuss the specific set of normalized WPH coefficients used in this work. In Appendix~\ref{app:sim-results}, we present the validation of the component-separation method on a set of \Planck simulations.
\begin{center}
\begin{table*}[!htbp]
\centering
\caption{Details of the sky regions used in our analysis: region number, centre (Galactic coordinates), \SNR of the dust signal, $\langle \NHI \rangle$, $99\%$ central percentile interval of \NHI,  the Pearson correlation coefficient between the data $m$ and \NHI, and $\epsilon_{353}$. Patch size of $14.9\degree \times 14.9\degree$ around the centre Galactic coordinates are cut from the full-sky \healpix map at $\Nside=512$. }
\label{tab:region-info}
\begin{tabular}{cc c c c c c}
\hline\hline 
\noalign{\vskip 2pt}
Region
& Galactic coordinates
& $\SNR$ 
& $\langle \NHI \rangle$
& 99\% central percentile \NHI 
& $\rho$ 
& $\epsilon_{353}$\\
\cline{2-7}\noalign{\vskip 2pt}

& $\paren{l,b}~[\deg]$
&
& $[\hiunit]$
& $[\hiunit]$
&
& $[\kJysr (\hiunit)^{-1}]$\\
\noalign{\vskip 2pt}\hline 

$\rm R_{1}$ & $(337.5\degree,-66.4\degree)$ 
& 0.9
& 1.2
& $0.8-2.1$  
& 0.38
& $33.33\pm 0.03$\\

$\rm R_{2}$ & $(15.0\degree,54.3\degree)$ 
& 3.0
& 2.5
& $1.3-4.1$
& 0.91
& $42.20\pm 0.01$\\

$\rm R_{3}$ & $(330.0\degree,-40.0\degree)$ 
& 4.6
& 3.4
& $1.8-6.8$
& 0.92
& $36.95\pm 0.01$\\
\hline\hline
\end{tabular}
\end{table*}
\end{center}
\section{Data used in the analysis}\label{sec:data-set}
\subsection{Datasets}\label{sec:data}
We use the \Planck Public Release 3 spectral matching independent component analysis (\texttt{SMICA}) CMB-subtracted intensity map at $353 \GHz$ from the \Planck legacy archive\footnote{\href{http://pla.esac.esa.int/pla}{http://pla.esac.esa.int/pla}}~\citep{planck-I:2018} for our analysis. The map is originally expressed in unit of $\kcmb$ and projected on a \healpix\footnote{\href{http://healpix.sf.net}{http://healpix.sf.net}}~\citep{Gorski:2005} grid at $\Nside=2048$ (pixel size $1.7\arcm$) with an angular beam resolution of $4.82\arcm$ full width at half maximum (FWHM)~\citep{planck-III:2018, planck-IV:2018}. We convert the unit of the map from $\kcmb$ to $\kJysr$ using the factor given in~\citet{planck-IX:2014}. We choose to work with CMB-subtracted intensity map ($m$) at $353\,\GHz$, because the two main diffuse foreground components present at high Galactic latitudes are the Galactic dust emission and the CIB anisotropies. Other foreground components such as free-free, synchrotron, and anomalous microwave emission are subdominant at this frequency. We avoid those line of sights where the bright infrared point sources are present at high Galactic latitudes. We remove the global offset of $119.3\,\kJysr$ estimated from the template-fit approach in~\citet{sinha2026aa}. The global offset term accounts for the contribution of the CIB monopole~\citep{B_thermin:2012} and other emission component that is not accounted for by the \HI emission. 

We use the low-velocity (LV) and intermediate-velocity (IV) components of the full-sky spectroscopic \HI data from the \texttt{HI4PI}\footnote{\href{http://cdsarc.u-strasbg.fr/viz-bin/qcat?J/A+A/594/A116}{http://cdsarc.u-strasbg.fr/viz-bin/qcat?J/A+A/594/A116}}\citep{hi4pi2016aa} to trace the dust emission at high Galactic latitudes. The LV \HI column density (\NLV) is integrated over the  velocity range $\abs{v_{LSR}} < 30\,\kms$, where $v_{LSR}$ is the velocity defined with respect to the local standard of rest. Similarly, the IV \HI column density (\NIV) is integrated over the  velocity range $30\,\kms < \abs{v_{LSR}} < 100\,\kms$. The two \NLV and \NIV maps are expressed in units of $10^{18}\,\cm^{-2}$ and have an angular beam resolution of $16.2\arcm$ FWHM projected on the \healpix grid at $\Nside=1024$ (pixel size $3.4\arcm$). We add the two \NLV and \NIV maps as a measure of the total \HI column density map (\NHI).

We use the all-sky Galactic dust reddening map\footnote{\href{https://doi.org/10.5281/zenodo.8207175}{https://doi.org/10.5281/zenodo.8207175}} (in units of $\rm mag$) produced by~\cite{chiang2023apj}, hereafter the CSFD map ($\scfd$), at a beam resolution of $6.1\arcm$ FWHM and the \healpix pixel resolution of $\Nside=2048$. The CIB-cleaned CSFD map at $100\,\mum$ is derived from the original~\citet*[][SFD]{SFD:1998} dust extinction map. In our analysis, we compare the \Planck component-separated dust map at $353\,\GHz$ with the  $\scfd$ map.
\subsection{Sky regions selection}\label{sec:sky-patch}

First, we smooth both $m$ and $\scfd$ maps to a common  beam resolution of $16.2\arcmin$ FWHM, which is the same as that of the \HI column density maps, and then downgrade all the maps to \healpix  resolution of $\Nside=512$ (pixel size $6.8\arcm$). We subsequently extract 2D square patches of area $222\deg^{2}$ (each side $14.9\degree$) with $256 \times 256$ pixels and a pixel size of $3.5\arcm$, centered on a \healpix pixel of $\Nside=4$, using the \texttt{reproject} python package~\citep{robitaille2020}. We select three sky regions with \SNR values ranging from $0.9$ to $4.6$, denoted as \Rlow, \Rint, and \Rhigh, respectively. Their sky positions in Galactic coordinates are listed in Table~\ref{tab:region-info}, together with \HI column-density statistics (mean; $\langle \NHI \rangle$ and $95\%$ percentile range) and the Pearson correlation coefficient, defined as $\rho$, between $m$ and the \NHI map. As shown in Table~\ref{tab:region-info}, $\rho$ increases with \SNR, indicating a progressively stronger correlation between the two maps. We model the total observed data as $m = \epsilon_{353}\,\NHI$, assuming a total noise level (CIB anisotropies and instrumental noise) of $9\,\kJysr$ (see Sect.~\ref{sec:contamination}) and $\epsilon_{353}$ to be spatially uniform across the entire region. The corresponding best-fit values of $\epsilon_{353}$ are reported in Table~\ref{tab:region-info}. The inferred values of $\epsilon_{353}$ for the three sky regions are consistent with those reported in previous studies of high Galactic latitude dust emission~\citep{planck-XVII:2014, T_Ghosh:2017, Adak:2019}. These regions thus provide distinct ISM environments, serving a useful framework for testing the component-separation algorithm and for studying Galactic dust emission at high Galactic latitudes and across a range of \HI column densities. 
\subsection{Synthetic contamination maps}\label{sec:contamination}

We use the $100$ statistically identical synthetic contamination realisations ($\rsyn$), constructed using the maximum entropy generative models, as the noise training set in the component-separation algorithm. The details of the synthesis of the $\rsyn$ maps at $353\,\GHz$ using the SC statistics~\citep{cheng2024pnas, mousset2024aa} is discussed in details in~\citetalias[][Section 5.1]{sinha2026aa}. These $\rsyn$ maps follow the Gaussian distribution and have a mean standard deviation $\sigma_{\rsyn}$ of $9.0\,\kJysr$ and the estimated $1\sigma$ variation in the $\sigma_{\rsyn}$ is $0.4\,\kJysr$. 
\subsection{Template-fit maps}\label{sec:hmc-dust}

\begin{figure}[!htbp]
\centering
\includegraphics[width=\columnwidth, keepaspectratio=True]{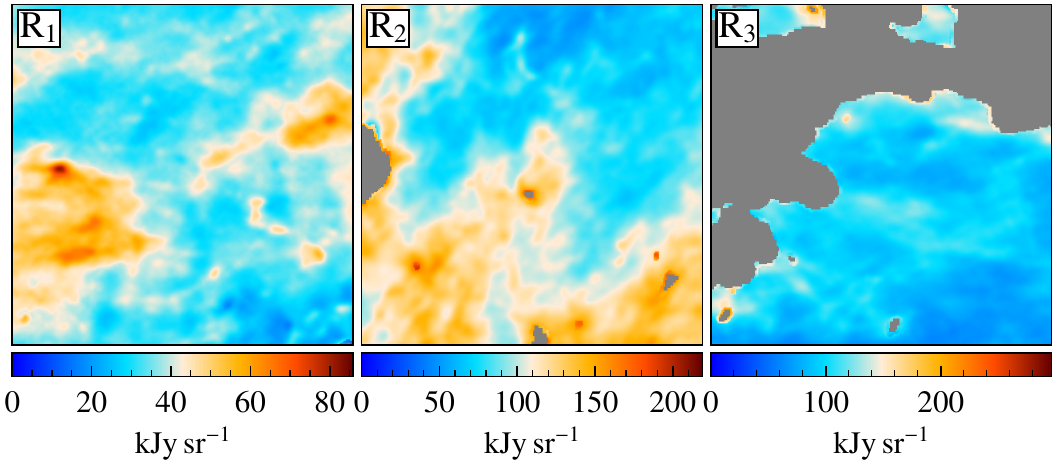}
\caption{Component-separated dust maps obtained from the template-fit approach for the three regions $\Rlow$, $\Rint$, and $\Rhigh$ (\textit{left} to \textit{right}).}
\label{fig:data-result-hmc-dust}
\end{figure}

We use the tight correlation between the dust emission and \HI at low dust column density regions (${\NHI < 4 \times \hiunit}$) within a template-fitting framework to extract the \HI-correlated dust emission~\citep{Boulanger:1996, planck-XVII:2014}. The pixels where the dust-\HI correlation breaks down are excluded from the subsequent analysis. Two \HI templates (\NLV and \NIV) are fitted simultaneously to the $m$ map over $1.8\degree \times 1.8\degree$ patch size (corresponding to $\Nside=32$ pixel size), together with a constant global offset, using the Hamiltonian Monte Carlo algorithm described in \cite{adak2024mnras}. The fitted global offset is subtracted from the resulting \HI-correlated dust emission map to produce a reference dust map, denoted by $\sref$. The $\sref$ map has an angular resolution of $16.2\arcmin$ FWHM. After applying the template-fitting mask with the \NHI threshold ${\NHI > 4 \times \hiunit}$, the percentage of sky available in \Rlow, \Rint, and \Rhigh are $99.9\%$, $98.5\%$, and $61.5\%$, respectively. The $\sref$ maps for \Rlow, \Rint and \Rhigh are shown in Fig.~\ref{fig:data-result-hmc-dust} from left to right, respectively. For each region, we also construct a reference contamination map ($\rref=m-\sref$), by subtracting the corresponding $\sref$ map from the total data map $m$. Since the template-fitting method recovers the dust emission over $99.9\%$ of the \Rlow region, the resulting $\sref$ map serves as a reliable independent reference to assess the robustness of the dust reconstruction obtained with the WPH-based component-separation method. Furthermore, the corresponding $\rref$ map is also used to generate 4 of the $100\,\rsyn$ realisations used in our analysis.
\section{component-separation algorithm}\label{sec:wph-algorithm}

In this section, we briefly outline the WPH-based component-separation algorithm used to decompose the data map $m$ into a dust signal map ($\ts$) and a contamination map ($\tr = m - \ts$). The framework is based on minimising a loss function defined over an ensemble of statistical constraints constructed directly from the data, as presented in~\cite{auclair2024aa, tsouros2026ar}. The constraints closely follow those introduced in~\citetalias{sinha2026aa} and are summarized below as,
\begin{align}
        &\Phi_{k}\paren*{m,\ts} \simeq \innerp*{ \Phi_{k}\paren*{ \ts+ \rsyni, \ts} }_{i\in N},\label{eq:constaint-1}\\
        &\Phi_{k}\paren*{m - \ts} \simeq \innerp*{\Phi_{k}\paren*{\rsyni}}_{i\in N}\label{eq:constaint-2}, \\
        &\Phi_{k}\paren*{\NHI, m} \simeq \innerp*{\Phi_{k}\paren*{\NHI, \ts + \rsyni} }_{i\in N},\label{eq:constaint-3}\\
        &\Phi_{k}\paren*{\NHI, m-\ts} \simeq \innerp*{\Phi\paren*{\NHI, \rsyni}}_{i\in N}, \label{eq:constaint-4}
\end{align}
In these equations, $\Phi_{k}$ is the set of ${k^{th}}$ coefficients of the auto- and cross- normalised WPH and scaling moments (detailed in Appendix~\ref{app:wph-coeff}), $\rsyni$ is the ${i^{th}}$ realisation of the synthesised contamination map and $\innerp{\cdot}_{i}$ is the ensemble average over the $N$ realisations of $\rsyn$. The first constraint, Eq.~(\ref{eq:constaint-1}), enforce the statistical compatibility of $\ts$ and $\tr$ with the observed map $m$. The second constraint, Eq.~(\ref{eq:constaint-2}), leverages that the residual $m-\ts$ follows the same statistical distribution as the expected contamination $\rsyn$. The third constraint, Eq.~(\ref{eq:constaint-3}), exploits the expected statistical dependence between $\ts$ and the \NHI map, while the fourth constraint, Eq.~(\ref{eq:constaint-4}), imposes that residual $m-\ts$ is statistically uncorrelated with the \NHI map.
The loss functions corresponding to these constraints are:
\begin{eqnarray}
    \begin{aligned}\label{eq:loss-part2}
&\cL_{1}\paren*{u} = \frac{1}{N N_{k}}\sum_{i=1}^{N}\sum_{k}^{N_{k}}\loss*{\Phi_{k}\paren*{m,u} - \Phi_{k}\paren*{u + \rsyni, u}}^{2},\\
&\cL_{2}\paren*{u} = \frac{1}{N_{k}}\sum_{k}^{N_{k}}\loss*{\Phi_{k}(m-u) - \innerp*{\Phi_{k}\paren*{\rsyni}}_{i\in N}}^{2}, \\
&\cL_{3}\paren*{u} = \frac{1}{N N_{k}}\sum_{i=1}^{N}\sum_{k}^{N_{k}}\loss*{\Phi_{k}\paren*{\NHI, m} - \Phi_{k}\paren*{\NHI, u + \rsyni}}^{2},\\
&\cL_{4}\paren*{u} = \frac{1}{N_{k}}\sum_{k}^{N_{k}}\loss*{\Phi_{k}\paren*{\NHI, m-u} - \innerp*{\Phi_{k}\paren*{\NHI, \rsyni}}_{i\in N}}^{2},
\end{aligned}
\end{eqnarray}
where, $\loss{\,\cdot\,}$ is the Euclidean norm. The total loss function is given as
\begin{equation}
\cL\paren*{u} = \cL_{1}\paren*{u} + \cL_{2}\paren*{u} + \cL_{3}\paren*{u}+ \cL_{4}\paren*{u} .\label{eq:loss-total}
\end{equation}
\begin{figure*}[!htbp]
\centering
\includegraphics[width=0.8\linewidth, keepaspectratio=True]{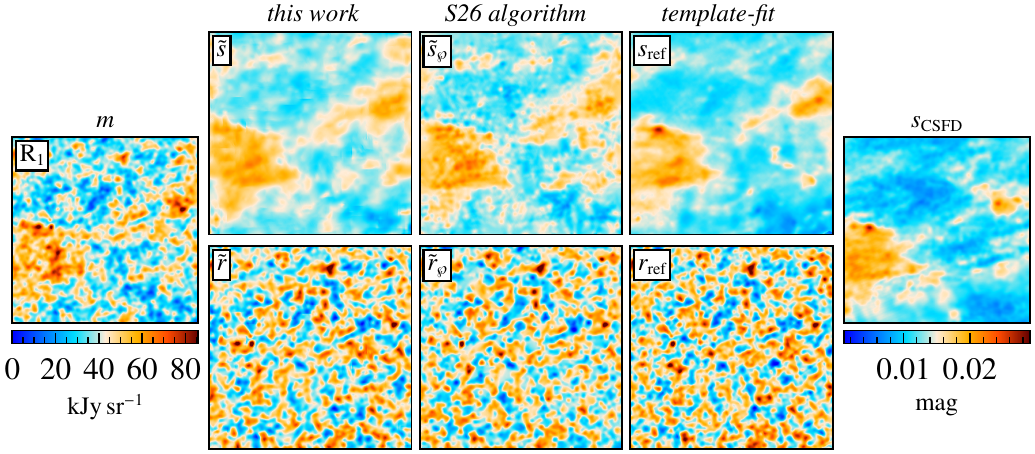}
\caption{The first column shows the \Planck $353 \GHz$ map (CMB and CIB monopole subtracted; $m$; \textit{left}) in \kJysr for \Rlow. The subsequent $3\times2$ grid shows the maps recovered from the different component-separation algorithms --- \textit{upper panel} shows the recovered dust maps and the \textit{lower panel} shows the recovered contamination maps. The dust maps use the same colour scale as the corresponding $m$ map while the contamination maps are shown over the range $\pm 30\,\kJysr$. The last column shows the dust reddening map at $100\,\mum$ in $\rm mag$.}
\label{fig:data-result-map-R1}
\end{figure*}     

Starting from an initial map $u_{0}$, we recover the signal by minimizing the total loss in Eq.~(\ref{eq:loss-total}) with a gradient-based optimization scheme in pixel space, specifically the limited-memory Broyden–Fletcher–Goldfarb–Shanno (L–BFGS) algorithm implemented in \texttt{PyTorch}~\citep{nocedal1980mc}, with the running map $u$ as the optimization variable. We smooth the $m$ map with a Gaussian beam of FWHM $100\arcm$ and use it as the $u_0$ map. We normalise the WPH coefficients using the $\sref$ maps, discussed in Sect.~\ref{sec:hmc-dust}. In \Rlow, approximately $99.9\%$ of the pixels are available in $\sref$ map, the remaining few masked pixels are inpainted using linear interpolation from the \texttt{scipy.interpolate} package~\citep{virtanen2020}. In \Rint and \Rhigh, the larger number of pixels are masked owing to the high \HI column densities; the missing pixels are filled using the corresponding values from the $m$ maps. Upon convergence, the final map $u$ is identified as the recovered signal map $\ts$. The optimisation is stopped after 80 steps or when the difference between the two consecutive losses is less than $10^{-3}$. At each iteration, $\cL_{1}$ and $\cL_{3}$ are computed over $N=100\,\rsyn$ maps to estimate the expectation over the contamination. Using $2477$ auto and $5108$ cross coefficients, each optimization run takes roughly around 2 hours on a single-node \texttt{NVIDIA A30} GPU, depending on the number of iterations performed.

In the implementation described in this paper, the losses used to minimise the differences between the terms being equated in Eq.~(\ref{eq:constaint-1}) to Eq.~(\ref{eq:constaint-4}) are estimated at each iteration based on the running value $u$ of the estimate of $s$. In particular, this means that the biases in the statistics caused by noise in Eq.~(\ref{eq:constaint-1}) to Eq.~(\ref{eq:constaint-4}) are re-evaluated at each iteration. In contrast, the implementation of~\citetalias{sinha2026aa} estimates these biases once and for all at the beginning of the optimisation. However, this latter approach can lead to a flawed estimate of the experimental biases, which can depend strongly on the map on which they are estimated, particularly when the \SNR ratio of the data is low\footnote{When noise is added to an already highly noisy signal rather than to a structured signal with little noise, it changes its non-Gaussian structure less.}. Furthermore, in the current paper, the loss terms are normalised relative to the signal intensity of a reference map. By contrast, the implementation of~\citetalias{sinha2026aa} was carried out using a $\chi^2$-based form, which tends to place a high weighting on highly noisy scales, making its optimisation more susceptible to the aforementioned problem. We believe that these two effects combine to explain the difference in results observed in this paper. For more details on the two approaches, see~\citet{delouis2022aa, auclair2024aa}.

We validate our component-separation algorithm on mock \Planck data for $\SNR=1$ case, as presented in Appendix~\ref{app:sim-results}.

\section{\Planck results }\label{sec:data-results}
We apply our component-separation algorithm to extract the dust signal from the \Planck $353\,\GHz$ data in all three sky regions. A key advantage of the present approach  over the component-separation algorithm adopted in~\citetalias{sinha2026aa} that it can recover the dust signal not only in the high $\SNR$ regions like \Rint and \Rhigh, but also in low \SNR regions such as \Rlow. Also, unlike the template-fit approach we do not rely solely on dust-\HI correlation to separate the dust signal. Hence, heavily masked region $\Rhigh$ of template-fit approach can be easily handled in our methodology. The \Rlow region serves as a benchmark for validating the reconstruction of $\ts$, due to the availability of $\sref$ obtained in Sect.~\ref{sec:hmc-dust}. We also compare the $\ts$ map in \Rlow with the dust map ($\tssc$) derived using \citetalias{sinha2026aa} algorithm.
\subsection{Recovered dust maps}\label{sec:data-wph-maps}

We present the component-separated maps for region $\Rlow$ in Fig.~\ref{fig:data-result-map-R1}. The first column shows the data map $m$, while the subsequent $3\times2$ grid shows the maps recovered from the different component-separation algorithms. The upper panel of the grid shows the extracted dust signals from the three methods, while the lower panel shows the corresponding contamination maps. The absence of small-scale CIB anisotropies in the $\ts$ map in the ``\textit{this work}'' column confirms that the current component-separation algorithm successfully recovers the dust signal buried underneath the contamination. The $\tssc$ map in the ``\textit{S26 algorithm}'' column clearly shows the grainy features at small angular scales, whereas the $\sref$ from the ``\textit{template-fit}'' column is much smoother. The excess small-scale power in $\tssc$ map arises from the loss-minimisation approach adopted in~\citetalias{sinha2026aa}, as explained in Sect.~\ref{sec:wph-algorithm}. On the other hand, the $\sref$ map lacks structures below $1.8\degree$, as the dust emissivity variations are estimated only at that scale (see Sect.~\ref{sec:hmc-dust}). Visually, the phase information in the contamination maps agrees well across all three methods. The differences between the three recovered dust maps are further quantified in terms of angular power spectrum in Appendix~\ref{app:data-method-comp}. The last column shows the reddening map $\scfd$ at $100\,\mum$. At large angular scales, the $\ts$ map correlate well with the $\scfd$ map, but there are map-level differences at small angular scales.
\begin{figure}[!hbtp]
\centering
    \includegraphics[width=\columnwidth, keepaspectratio=True]{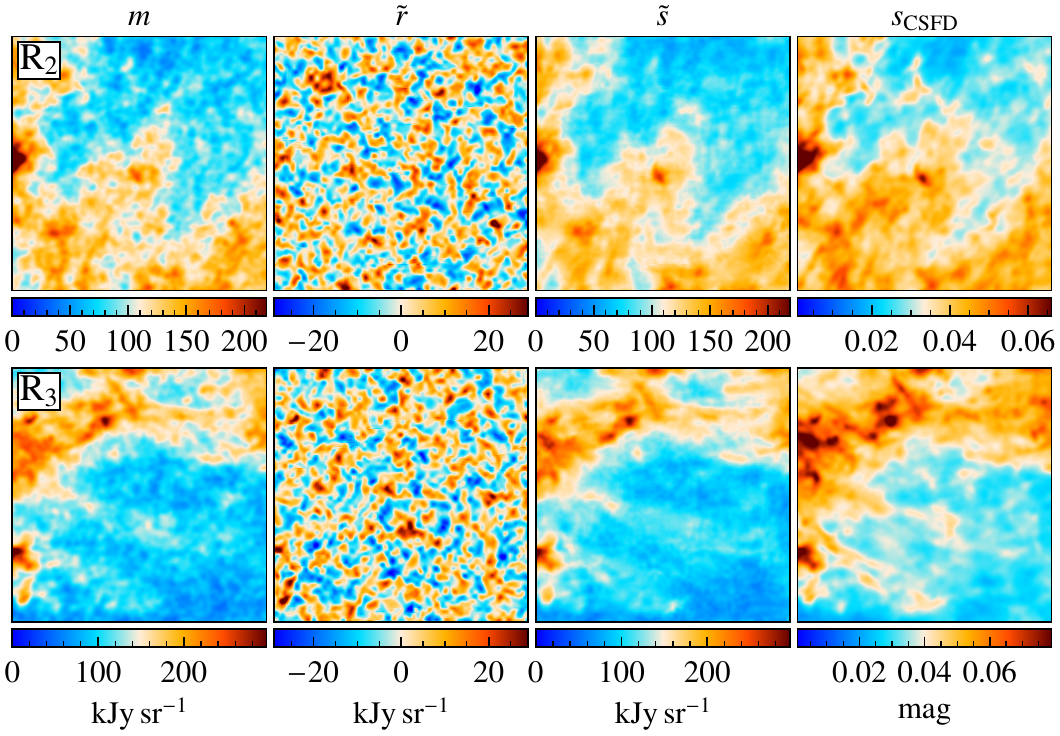}
\caption{ From \textit{left} to \textit{right}: the CMB-subtracted Planck map, the component-separated contamination map, the component-separated dust signal map, and the CSFD map at $100\,\mum$. The top panel shows the results of region $\Rint$, and the bottom panel shows the same for $\Rhigh$.} 
\label{fig:data-result-map-R23}
\end{figure}     

The upper and lower panels of Fig.~\ref{fig:data-result-map-R23} show the $m$, $\tr$ and $\ts$ maps for the \Rint and \Rhigh regions, respectively. The small scale fluctuations present in the $m$ maps are largely absent in the $\ts$ maps, which retain the large scale structures. The current formalism is also able to recover the dust signal within the masked regions (see Fig.~\ref{fig:data-result-hmc-dust}), as evidenced by the absence of significant dust residuals in the corresponding $\tr$ maps. The last column shows the $\scfd$ maps for the \Rint and \Rhigh regions, respectively. In both regions, the recovered $\ts$ maps are in good agreement with the $\scfd$ maps, at large angular scales, although noticeable differences remain at smaller scales.
\begin{figure*}[!hbtp]
\centering
\includegraphics[width=\linewidth]{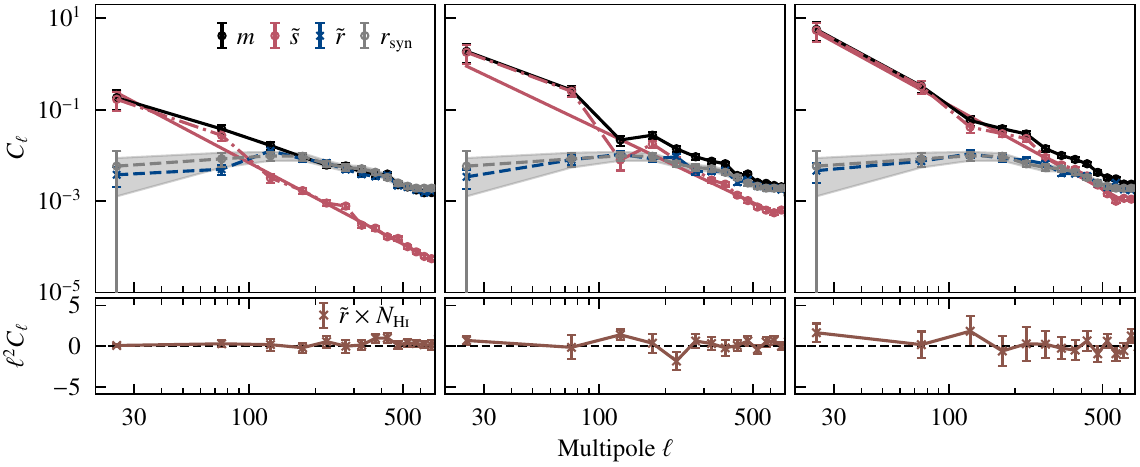}
\caption{\textit{Upper panel}: Auto power spectra $\Cl$ of $m$ (black solid line), $\ts$ (red dot-dashed line), $\tr$ (blue dashed line) and the mean and standard deviation of $100$ realisations of $\rsyn$ (gray dashed line). The gray band shows the 16 and 84 percentiles for the $\rsyn$ maps. The red solid line shows the best-fit power-law model for $\ts$. The three columns are for the \Rlow, \Rint and \Rhigh regions, respectively. \textit{Lower panel}: Cross power spectra of $\tr$ with \NHI in $\ell^{2}\Cl$ for the three respective regions in the unit $\kJysrsq$.}
\label{fig:data-result-ps-cl} 
\end{figure*}
\subsection{Power spectrum analysis}

Figure~\ref{fig:data-result-ps-cl} shows the angular power spectra ($\Cl$) with multipoles $\ell$ of the three regions, \Rlow, \Rint, and \Rhigh (from left to right). We compute the $\Cl$ using \namaster~\citep{alonso2019mnras}, adopting the discrete Fourier transform in the flat sky approximation~\citep{Hivon:2002, ponthieu2011aa}. To reduce boundary effects, we apply a cosine-apodised mask to the central $180\times180$ pixels~\citepalias[see Fig.~B.1 of][]{sinha2026aa}. The resulting spectra are corrected for beam and pixel-window effects. We estimate the covariance matrix of the binned angular power spectra using the Gaussian covariance estimator implemented in \namaster~\citep{Tristram:2006, alonso2019mnras}. This estimator assumes that the underlying fields are Gaussian and computes the expected covariance from the corrected power spectra. The resulting covariance matrix quantifies both the uncertainties of the measured bandpowers and the correlations between different multipole bins. We use the square root of its diagonal elements as the error bars ($\sigma_{\Cl}$) on the measured power spectra in all subsequent statistical analyses. The black solid, red dot-dashed, and blue dashed lines correspond to the auto-power spectra of $m$, $\ts$, and $\tr$, respectively. The crossover multipole at which the $\ts$ and $\tr$ power spectra intersect varies with the $\SNR$ of the region, occurring at $\ell = 100$, $225$, and $425$ for regions $\Rlow$, $\Rint$, and $\Rhigh$, respectively. This reflects the increasing dominance of the dust signal over the contamination at higher $\SNR$. For the high-$\SNR$ region $\Rhigh$, the $\ts$ power spectrum dominates or is comparable to the $\tr$ power spectrum across all angular scales up to the beam resolution of the \Planck $353\,\GHz$ frequency band. In contrast, in the low $\SNR$ region $\Rlow$, the $\tr$ power exceeds that of the $\ts$ across nearly all multipoles, with the exception of the first two multipole bins. The $\ts$ spectra are well described by a power-law model, $\Cl \propto (\ell/75)^{\alpha}$, fitted over the multipole range $25 \leq \ell \leq 625$. The best-fit models are shown by the solid red lines, with slopes $\alpha_{\ts}=-2.66\pm0.08$, $-2.33\pm0.17$, and $-2.66\pm0.12$ for \Rlow, \Rint, and \Rhigh, respectively. For comparison, we also compute the auto-power spectra of the \NHI and \NLV maps. In \Rlow, the best-fit slopes are $\alpha_{\HI}=-2.31\pm0.08$ and $\alpha_{\rm LV}=-2.60\pm0.08$, indicating that the recovered dust emission is strongly correlated with the \NLV component. In contrast, the corresponding \NHI and \NLV power spectra in the \Rint and \Rhigh regions have slopes of $\alpha_{\HI}=-2.76\pm0.15$ and $-2.93\pm0.08$, and $\alpha_{\rm LV}=-2.78\pm0.16$ and $-2.96\pm0.07$, respectively. The significant difference between these slopes and that of the recovered dust spectrum indicates that the tight dust-\HI correlation gradually breaks down in the higher \HI column-density regions. This behaviour is consistent with the presence of an additional dust component associated with molecular hydrogen (\molH), as demonstrated in~\citetalias{sinha2026aa}, as well as spatial variations in the dust emissivity, as discussed in Sect.~\ref{sec:dust-hi}.

We further compare the power spectra of the recovered contamination maps with those of the 100 $\rsyn$ maps. The gray dashed line represents the mean spectrum of the realisations, while the gray band encloses the 16 and 84 percentile range. For all three regions, the recovered $\tr$ spectra are in excellent agreement with the expected spectra from the $\rsyn$ ensemble, indicating that the contamination power is preserved at small angular scales and that the dust signal has been successfully separated from the contamination. Finally, we compute the cross-power spectrum between $\tr$ and \NHI to test for residual dust leakage in the recovered contamination maps. The lower panels of Fig.~\ref{fig:data-result-ps-cl} show the $\tr \times \NHI$ cross-spectrum, plotted in $\ell^{2} \Cl$. Within the $1\sigma$ uncertainties derived from the Gaussian approximation, no statistically significant correlation is detected for any of the three regions, indicating negligible leakage of dust emission into the recovered contamination maps.
\subsection{Non-Gaussianity test of $\tr$ maps}

\begin{figure}[!htbt]
\centering
\includegraphics[width=\columnwidth]{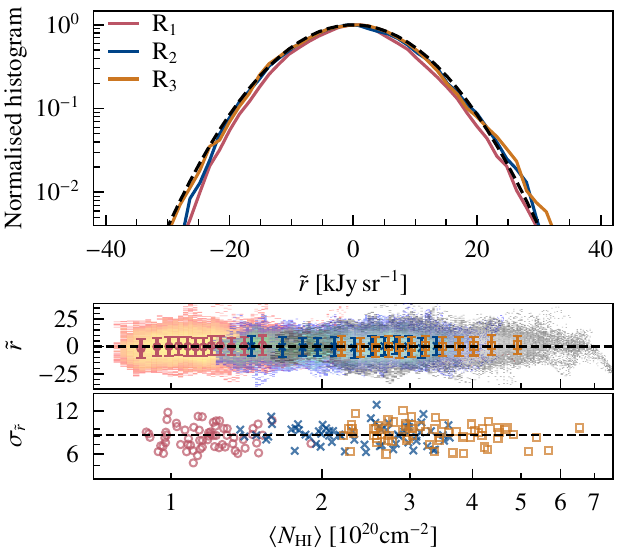}
\caption{\textit{Upper panel}: One dimensional normalised distributions of $\tr$ (in solid lines) of $\Rlow, \Rint$ and $\Rhigh$. The black dashed line shows the best-fit Gaussian model with a standard deviation of $\sigma_{\rsyn}=9.0\,\kJysr$. \textit{Middle panel}: Shows the non-correlation of $\tr$ with \NHI map for the three regions. The data points and the errorbars show the zero median and standard deviation of $\tr$ in the ordered bins of \NHI. The standard deviations are consistent in all the \NHI bins. \textit{Lower panel}: Shows the non-correlation of the standard deviation of $\sigma_{\tr}$ with mean \NHI calculated within $32\times32$ sub-pixels. The black dashed line represents the median of $\sigma_{\tr}$.} 
\label{fig:data-result-cib-hist} 
\end{figure}
We compute the normalised histogram of the $\tr$ maps in the three regions. The normalised histograms are presented in the upper panel of Fig.~\ref{fig:data-result-cib-hist}, along with the mean expected Gaussian distribution of $\sigma_{\rsyn}=9\,\kJysr$ derived from 100 $\rsyn$ maps. From the observed distribution, it is clear that the statistics of three $\tr$ maps follows closely the Gaussian distribution. We fit a Gaussian to the observed distribution of $\tr$ and find $1\sigma$ standard deviation of $8.1, 8.8$, and $8.8\,\kJysr$ for region $\Rlow$, $\Rint$ and $\Rhigh$, respectively. These values are consistent with the $1\sigma$ statistical variance of $0.4$\,\kJysr derived from the $100\,\rsyn$ maps. The middle panel of Fig.~\ref{fig:data-result-cib-hist} shows the combined 2D scatter plot of $\tr$ with \NHI for the three sky regions. We bin the $\tr$ map per sky region into 10 bins in order of increasing value of \NHI and compute the median values along with the 16 and 84 percentiles of the data over each bin. We plot the median values of $\tr$ and the percentiles as upper and lower limits of the error bars in  Fig.~\ref{fig:data-result-cib-hist}. The $1\sigma$ error bar on $\tr$ remains almost the same as we probe different ISM environments from $0.8\times \hiunit$ to $6.8\times\hiunit$. There is no systematic dependence seen between the $\tr$ map and \NHI at the pixel level.  
In the lower panel of Fig.~\ref{fig:data-result-cib-hist}, we present the variation of the standard deviation of $\tr$ ($\sigma_{\tr}$) with the mean \NHI, computed within $32\times 32$ pixels, for the $\Rlow$, $\Rint$, and $\Rhigh$ regions. The black dashed line represents the combined median value from the three regions, $\sigma_{\tr}=8.2\,\kJysr$, while the standard deviation is $1.4\,\kJysr$. As shown in the figure, $\sigma_{\tr}$ exhibits no statistically significant trend with the mean \NHI, implying that $\tr$ is effectively independent of \NHI. 
\begin{figure}[!htbp]
\centering
\includegraphics[width=\columnwidth]{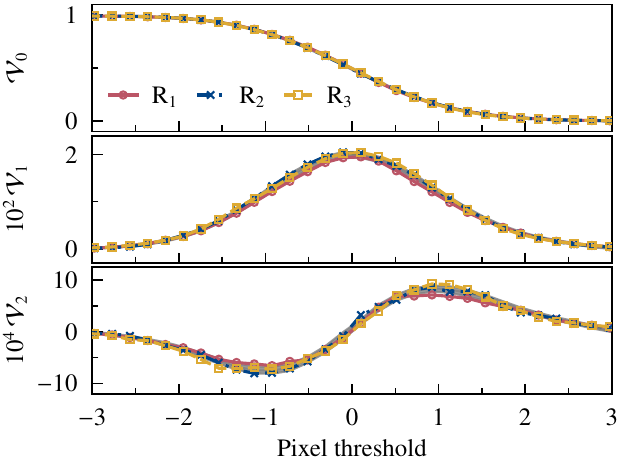}
\caption{The MFs of $\tr$ for \Rlow (red dots with solid line), \Rint (blue crosses with dashed-dot line) and \Rhigh (orange squares with dashed line). The solid gray lines show the MFs computed from the 100 $\rsyn$ realisations. We scale $\mathcal{V}_{1}$ and $\mathcal{V}_{2}$ with $10^{2}$ and $10^{4}$, respectively, for visualisation purpose.} 
\label{fig:app-result-wph-mkf} 
\end{figure}
Finally, we compute the scalar Minkowski Functionals (MFs) of mean-subtracted normalised $\tr$ maps to search for deviations from the statistical properties of the 100 $\rsyn$ maps. We use the publicly available \texttt{QuantImPy} package~\citep{mantz2008jsm,boelens2021} to compute the three scalar MFs: the area fraction ($\mathcal{V}_{0}$), perimeter ($\mathcal{V}_{1}$), and Euler characteristic ($\mathcal{V}_{2}$). Figure~\ref{fig:app-result-wph-mkf} shows the MFs of the $\tr$ maps for the three sky regions together with the corresponding distributions derived from the $100\,\rsyn$ realisations. We find good agreement between them, indicating that the recovered contamination maps are statistically consistent with the expected contamination and show no evidence of residual dust leakage. For the low \SNR region \Rlow, any residual dust contribution in $\tr$ is expected to be small because the emission is contamination dominated. The \Rint and \Rhigh regions provide a more stringent test, as even a small leakage of dust signal into the contamination maps could produce detectable deviations in both the two-point statistics and the MFs. The consistency observed in these regions therefore provides additional evidence for the robustness of the component-separation procedure.

From our analysis, we concluded that the statistics of $\tr$ maps for the three regions follows closely the Gaussian distribution and is statistically isotropic.
\subsection{Dust-\HI correlation}\label{sec:dust-hi}

\begin{figure*}[!hbtp]
\centering
\includegraphics[width=0.8\linewidth, keepaspectratio=True]{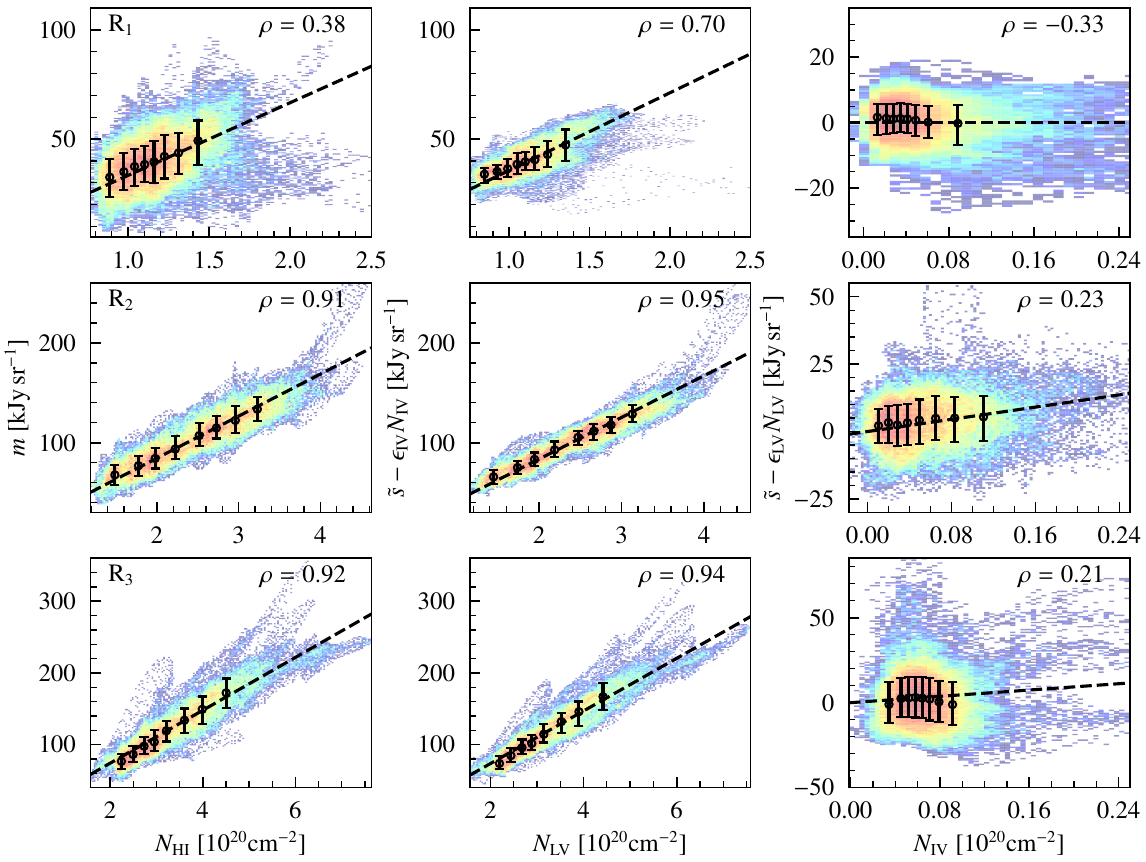}
\caption{Scatter plot of $m$ vs. $\NHI$ (\textit{left column}), $\ts$ vs. $\NLV$ (\textit{middle column}), and $\ts$ vs. $\NIV$ (\textit{right column}) to visualize the dust-\HI correlation for \Rlow, \Rint and \Rhigh from \textit{top to bottom}, respectively. The black circles correspond to the median values and the upper and lower limits correspond to the 16 and 84 percentile of the quantity represented in the $y$-axis in each ordered bin of the quantity represented in the $x$-axis. The best-fit dashed lines in the first column show the linear regression between $m$ and \NHI, while those in the second and third columns show the linear regressions of $\ts$ against \NLV and \NIV, respectively, from the two-template fit.}
\label{fig:data-result-dust-hi-corr}
\end{figure*}   

Figure~\ref{fig:data-result-dust-hi-corr} shows the two-dimensional histograms of $m$ with \NHI (first column), $\ts$ versus $\NLV$ (second column), and $\ts$ versus $\NIV$ (third column). The rows correspond to the \Rlow, \Rint and \Rhigh regions, from top to bottom, respectively. 
The black circles show the median values calculated for bins containing an equal number of pixels with increasing column densities. The associated error bars, represent the 16 and 84 percentiles of the distribution within each bin.
In the \Rlow region, the distribution of $m$ has a larger scatter with \NHI, primarily due to the contamination. This scatter is substantially reduced in the distribution of $\ts$ with \NLV, demonstrating that the algorithm effectively removes the contamination while preserving the dust signal. A similar reduction in scatter is seen for the \Rint and \Rhigh regions. 

The dashed lines in Fig.~\ref{fig:data-result-dust-hi-corr} show the results of linear regressions. For $\ts$, we consider distinct emissivities for low- and intermediate-velocity gas. 
\begin{equation}
    \ts = \epsilon_{\rm LV}\, \NLV  +  \epsilon_{\rm IV}\, \NIV \ ,\label{eq:dust-HI-fit0}
\end{equation}
The best-fit values of $\epsilon_{\rm LV}$ are $35.6 \pm 0.5$, $41.8 \pm 0.4$, and $36.7 \pm 0.4\,\kJysr\,(\hiunit)^{-1}$ for the \Rlow, \Rint, and \Rhigh regions, respectively. The corresponding best-fit values\footnote{We impose a positive prior on the coefficients of the \NIV template} of $\epsilon_{\rm IV}$ are $0 \pm 2.7$, $56.8 \pm 4.8$, and $46.4 \pm 4.8 \,\kJysr (\hiunit)^{-1}$, respectively. The inferred emissivities are broadly consistent with previous measurements of \HI-correlated dust emissivities at high Galactic latitudes \citep{planck-XXIV:2011,planck-XVII:2014, adak2024mnras}.

The Pearson correlation coefficient, $\rho$, shows that the $\ts$ map is more strongly correlated with the \NLV map than with the \NIV map in all three regions. In a very low \HI column density region $\Rlow$, the correlation between $\ts$ and \NIV is mildly negative.
In contrast, a weak but noticeable correlation between $\ts$ and \NIV is observed in region $\Rint$ and $\Rhigh$ when we probe relatively high \HI column density regions. 

\begin{figure}[!htbp]
\centering
\includegraphics[width=\columnwidth, keepaspectratio=True]{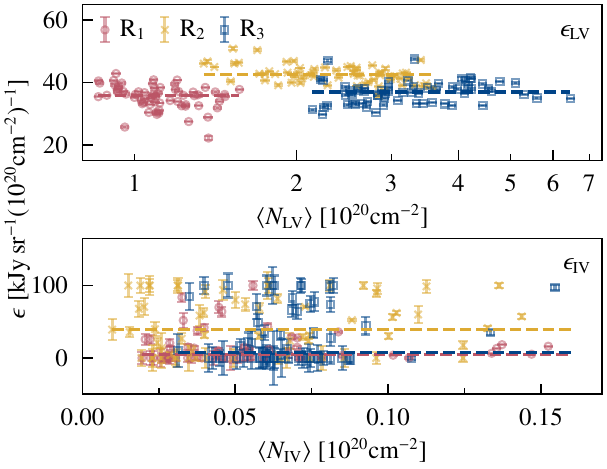}
\caption{Variation of $\epsilon_{\rm LV}$ (\textit{upper panel}) and $\epsilon_{\rm IV}$ (\textit{lower panel}) with the mean \NLV and \NIV, respectively, over $\Theta=1.87\degree$ sub-regions (64 sub-regions) for the three regions. The dashed lines represent the respective median values. }
\label{fig:data-em-fit}
\end{figure}     
\section{Variations of dust emissivity }\label{sec:dust-emissivity}
In Sects.~\ref{sec:analysis_maps} and \ref{sec:residual_maps}, we gather evidence of spatial variations in dust emissivity, i.e. the dust emission per hydrogen. In Sect.~\ref{sec:cross_spectra}, we characterize their scale dependence.

\subsection{Spatial analysis of the dust-\HI correlation}\label{sec:analysis_maps}

We use the component-separated maps to investigate the scatter of $\ts$ from its expected value derived using a mean value of dust-\HI correlation ratio. We model $\ts$ as a linear combination of the two \HI velocity components with emissivities that are assumed to remain constant over a correlation scale $\Theta$,
\begin{equation}
    \ts\paren{\Theta} = \epsilon_{\rm LV}\paren{\Theta} \NLV  +  \epsilon_{\rm IV}\paren{\Theta}\NIV \ ,\label{eq:stilde-corr-fit}
\end{equation}
where $\Theta$ denotes the angular scale over which the emissivities $\epsilon_{\rm LV}$ and $\epsilon_{\rm IV}$ are assumed to be constant. We consider three correlation scales to capture the local emissivity variations. We divide the  maps into 64 sub-regions, each corresponding to $\Theta = 1.87\degree$ ($32 \times 32$ pixels), and fit the values of the two  emissivities in Eq.~\eqref{eq:stilde-corr-fit} for each patch. For $\Theta = 14.9\degree$, the data model is that of Eq.~\eqref{eq:dust-HI-fit0} introduced in Sect.~\ref{sec:dust-hi}.

Figure~\ref{fig:data-em-fit} shows the best-fit values of $\epsilon_{\rm LV}$ and $\epsilon_{\rm IV}$ as a function of the mean \NLV and \NIV, respectively, across the 64 sub-regions of size $1.87\degree$ for the three regions. The upper and lower panels show the results for $\epsilon_{\rm LV}$ and $\epsilon_{\rm IV}$, respectively. The dashed lines indicate the  median values of $\epsilon_{\rm LV}$, which are $35.9$, $42.7$, and $36.9\,\kJysr\,(\hiunit)^{-1}$ for the three regions. The corresponding median values of $\epsilon_{\rm IV}$ are $4.6$, $39.4$, and $7.4\,\kJysr\,(\hiunit)^{-1}$. The standard deviations of the LV coefficients in the three regions are $3.8$, $2.9$, and $3.5\,\kJysr,(\hiunit)^{-1}$, respectively, while the corresponding values for the IV coefficients are $21.3$, $38.4$, and $38.5\,\kJysr\,(\hiunit)^{-1}$.
\subsection{Residuals to the dust-\HI correlation}\label{sec:residual_maps}


\begin{figure}[!hbtp]
\centering
\includegraphics[width=0.8\columnwidth]{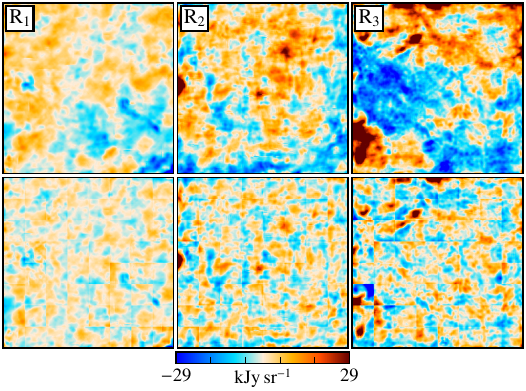}
\caption{Maps of residual emission, $\mathcal{E}$, computed for $\Theta$ equal to $14.9\degree$ (top) and $1.87\degree$ (bottom) for the three regions $R_1$ to $R_3$ from left to right.}
\label{fig:data-result-fit} 
\end{figure}
\begin{figure*}[!hbtp]
\centering
\includegraphics[width=0.7\linewidth]{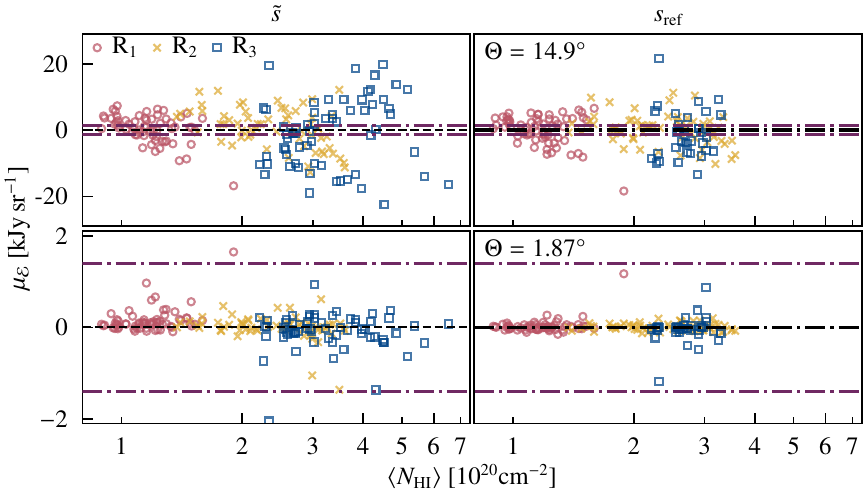}
\caption{Mean of the residual emission maps computed over sub-regions ($\muE$) vs  mean \HI column density for two  values of $\Theta$: $14.9\degree$ (\textit{top}) and $1.87\degree$ (\textit{bottom}). The plots for $\ts$ are on the left, and those for $\sref$ on the right. The black dashed line is the zero line and the two purple dot-dashed horizontal lines represents the corresponding $\pm1\sigma$ contamination levels, where $\sigma_{\rsyn}=1.4\,\kJysr$.} 
\label{fig:data-result-res-mean} 
\end{figure*}
\begin{figure}[!hbtp]
\centering
\includegraphics[width=\columnwidth]{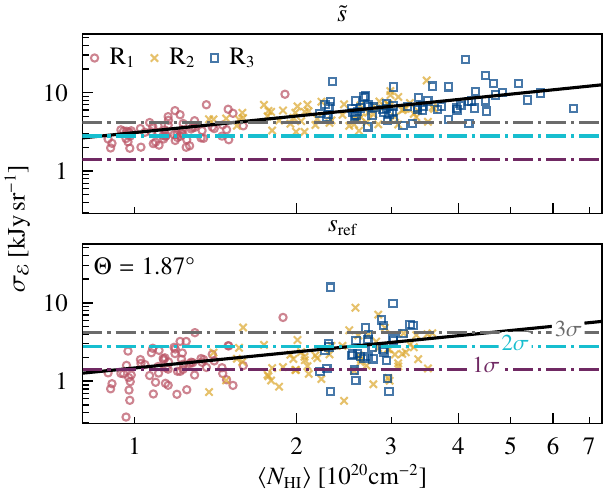}
\caption{Standard deviation of the residual emission maps computed over sub-regions ($\sigmaE$) vs the mean \HI column density for $\Theta=1.87\degree$. The top plot is for $\ts$ and the bottom plot for $\sref$. The black solid line is the best-fit model ($\sigmaE \propto \langle \NHI \rangle ^{\gamma}$), with $\gamma=0.7$ for both $\ts$, and $\sref$ maps. The purple, cyan, and gray dot-dashed horizontal lines indicate the corresponding $1\sigma$, $2\sigma$, and $3\sigma$ contamination levels, respectively.}
\label{fig:data-result-res} 
\end{figure}

To further characterize the spatial variations in the dust emissivity, we analyze maps of the residual emission, $\mathcal{E}\paren{\Theta}$,  obtained by subtracting the best-fit model of Eq.~\eqref{eq:stilde-corr-fit} from $\ts$. The two rows of Fig.~\ref{fig:data-result-fit} show the resulting $\mathcal{E}(\Theta)$ maps for $\Theta=14.9$ and $1.87\degree$. Variations of the residual emission is largest when a single emissivity is fitted over the entire patch ($\Theta=14.9\degree$) and decreases as the correlation scale becomes smaller, indicating that allowing for local dust emissivity variations significantly improves the residual map. To quantify this behaviour, we compute the mean ($\muE$) and standard deviation ($\sigmaE$) of the $\mathcal{E}$ map within each sub-region. 

The left column of Fig.~\ref{fig:data-result-res-mean} shows the variation of $\muE$ with $\langle \NHI \rangle$ for the three sky regions. The average value of $\muE$ remains consistent with zero when we combine the three regions for both correlation scales. The purple dot-dashed lines indicate the $\pm1\sigma$ contamination level, where $\sigma=1.4\,\kJysr$ is the standard deviation of $\sigma_{\tr}$ measured from the $\tr$ map (Fig.~\ref{fig:data-result-cib-hist}; lower panel). As $\Theta$ decreases, the scatter in $\muE$ around the mean expected value of zero decreases and becomes smaller than the $\pm1\sigma$ standard deviation of $\sigma_{\tr}$.

The upper panel of Fig.~\ref{fig:data-result-res} shows the dependence of $\sigmaE$ on the $\langle \NHI \rangle$ for the three sky regions. We fit the combined relation for the three regions with a power-law model, $\sigmaE \propto \langle \NHI \rangle^{\gamma}$, and determine the best-fit exponent $\gamma$ for each correlation scale. For $\Theta=1.87\degree$, we obtain $\gamma=0.7$. The corresponding best-fit models are shown by the black solid lines in Fig.~\ref{fig:data-result-res}. The purple, cyan, and gray dot-dashed horizontal lines indicate the corresponding $1\sigma$, $2\sigma$, and $3\sigma$ contamination levels, respectively. Except in the low column density \Rlow region, the increase in $\sigmaE$ with increasing $\langle \NHI \rangle$ exceeds the $3\sigma$ level, demonstrating statistically significant local variations in the dust emissivity. 

We perform a similar analysis of the $\sref$ maps for the three regions, using only the available pixels, which cover $98.5\%$ of $\Rint$ and $61.5\%$ of $\Rhigh$. We divide each region into 64 sub-regions and keep only those containing more than $20\%$ unmasked pixels before evaluating $\mathcal{E}$. We then calculate $\muE$ and $\sigmaE$ within the set of retained sub-regions. The  variations are displayed in the right column of Fig.~\ref{fig:data-result-res-mean} and in the lower panel of Fig.~\ref{fig:data-result-res}, respectively. For \Rlow and \Rint, the variation of $\sigmaE$ with the mean \NHI remains well below the $3\sigma$ level. In \Rhigh, however, approximately $24\%$ of the $\sigmaE$ measurements lie above the $3\sigma$ level and exhibit an increasing trend. This behaviour is expected because, in the template-fitting approach~\citepalias[][]{sinha2026aa}, the dust emissivity is assumed to be constant over each $1.8\degree\times 1.8\degree$ patch during the dust-\HI correlation analysis. Consequently, variations in the emissivity on scales smaller than $1.8 \degree$, as well as scatter in the dust-\HI correlation due to variations of the dust emissivity along the line of sight, are not captured in the $\sref$ maps.

The observed dependence of $\sigmaE$ on the mean \NHI shows that the small-scale dust-\HI decorrelation is not driven by $\tr$. This trend may arise from changes in the dust temperature and spectral index over the sky and along the lines of sight. This interpretation is consistent with the findings of the multi-frequency study of the dust-\HI correlation by \citet{planck-XVII:2014}. Emission from diffuse molecular and ionized gas could also contribute. 

\subsection{Scale dependence from cross-power spectra}\label{sec:cross_spectra}

To investigate the scale dependence of the dust-\HI correlation, we introduce the correlation ratio, defined as:
\begin{equation}
\mathcal{R}_{\ell} = \frac{C_{\ell}^{XY}} {\sqrt{C_{\ell}^{XX} \times C_{\ell}^{YY}}} \ ,
\end{equation}
where the indices $X$ and $Y$ represent $\ts$ and $\NHI$, respectively, and $C_{\ell}$ is the angular power spectrum. 
\begin{figure}[!htbp]
\centering
\includegraphics[width=\columnwidth, keepaspectratio=True]{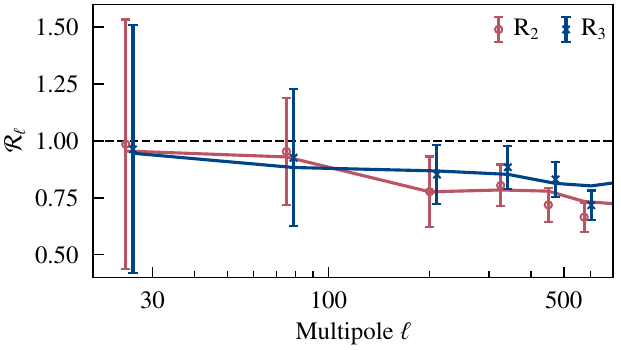}
\caption{Dust correlation ratio, $\mathcal{R}_{\ell}$, between $\ts$ and $\NHI$ for regions $\Rint$ (red) and $\Rhigh$ (blue). The mean $\mathcal{R}_{\ell}$ for 100 realisations of the modelled dust map with a spatially varying dust emissivity are shown in solid lines for regions $\Rint$ (red) and $\Rhigh$ (blue). The blue crosses are shifted slightly along the $x$-axis for clarity.}
\label{fig:data-result-corr-ratio}
\end{figure}     

The uncertainty on $\mathcal{R}_{\ell}$ is obtained by propagating the uncertainties of the corresponding auto and cross power spectra. We estimate the uncertainty as
\begin{equation}
\sigma_{\mathcal{R}_{\ell}} =
\mathcal{R}_{\ell}
\sqrt{
\paren*{\frac{\sigma_{C_{\ell}^{XY}}}{C_{\ell}^{XY}}}^2
+\frac{1}{4}\paren*{\frac{\sigma_{C_{\ell}^{XX}}}{C_{\ell}^{XX}}}^2
+\frac{1}{4}\paren*{\frac{\sigma_{C_{\ell}^{YY}}}{C_{\ell}^{YY}}}^2
},
\end{equation}
where $\sigma_{C_{\ell}^{XY}}$, $\sigma_{C_{\ell}^{XX}}$, and $\sigma_{C_{\ell}^{YY}}$ are the uncertainties in the cross power spectrum and the two auto power spectra, respectively.  Figure~\ref{fig:data-result-corr-ratio} shows the correlation coefficient, $\mathcal{R}_{\ell}$, as a function of $\ell$ for the  $\Rint$ and $\Rhigh$ regions. The uncertainty on $\mathcal{R}_{\ell}$ at the first multipole bin ($\ell=25$) is mostly dominated by the cosmic variance of the dust emission. The value of $\mathcal{R}_{\ell}$ being close to unity at small multipoles confirms a strong large-scale correlation between the dust and \HI emission. The gradual decline of $\mathcal{R}_{\ell}$ towards higher multipoles indicates that the dust and \HI emission become progressively decorrelated on smaller angular scales.

To explain the observed decorrelation ratio, we model the dust emission as $M = \epsilon_{\rm G} \NHI$, where $\epsilon_{\rm G}$ is a Gaussian realization of spatially varying dust emissivity following a power spectrum $\Cl \propto \ell^{\gamma}$. We obtain the mean and standard deviation of the dust emissivity by taking the simple ratio $q=\ts/\NHI$ for the two sky regions. We find the mean dust emissivity to be $\mu_{q} = 43.7$ and $36.6\,\kJysr\,(\hiunit)^{-1}$, with standard deviations $\sigma_q$ of $3.5$ and $4.0\,\kJysr,(\hiunit)^{-1}$ for \Rint\ and \Rhigh, respectively. We generate a full-sky emissivity map following the underlying power spectrum and extract a $14.9\degree \times 14.9\degree$ square patch around \Rint and \Rhigh. First, we subtract the mean of the 2D square patch and normalize it to have unit variance over the patch. We denote this map as $\epsilon_{\rm norm}$. We then scale the $\epsilon_{\rm norm}$ map to reproduce the observed emissivity mean and standard deviation for a given sky region using
\begin{equation}
\epsilon_{\rm G} = \mu_q + \epsilon_{\rm norm} \times \sigma_q \ .
\end{equation}
For our purpose, we fix $\sigma_q$ to $4\,\kJysr/\hiunit$ for both regions. We adopt power-spectrum slopes of $\gamma=-2.2$ and $-2.4$ for \Rint\ and \Rhigh, respectively, to reproduce the observed decorrelation ratio. We repeat this process 100 times to produce 100 different realizations of the emissivity map, fixing $\mu_q$, $\sigma_q$, and $\gamma$ for each region, and compute the model decorrelation ratio between $M$ and the \NHI maps. The mean model $\mathcal{R}_{\ell}$, obtained by averaging over the 100 realisations of the model dust map, is plotted as solid lines in Fig.~\ref{fig:data-result-corr-ratio}.

\section{Summary and Discussion}\label{sec:discussion}

We have successfully extracted the dust signal for three square patches at high Galactic latitude, each covering an area of $222\,\deg^{2}$, employing a WPH-based component separation method at the \Planck $353\,\GHz$ frequency. The algorithm exploits the statistical differences between dust, CIB, and noise signals, as well as the correlation between dust emission and \HI, through cross-WPH statistics. We summarize the main results of this study.
\begin{itemize}
\item The low-\SNR region, \Rlow, serves as a benchmark for reconstruction of dust signal because the reference dust map, $\sref$, obtained from the template-fitting approach, is available over nearly the entire field. The $\ts$ map is in excellent agreement with $\sref$ across all accessible scales, while the $\tr$ map reproduces the phase structure of the reference contamination map over the contamination dominated angular scales ($\ell > 100$). Unlike the template-fitting approach, which is limited by the fixed $1.8\degree$ correlation scale ($\Nside=32$), the WPH-based method is not tied to a predefined correlation scale and successfully recovers the dust emission structures on both large and small angular scales.
\item The WPH-based component-separation algorithm reconstructs component-separated maps in low \HI column-density regions ($\NHI < 6.8\times \hiunit$) at high Galactic latitudes. Unlike the standard template-fitting approach, the algorithm does not rely solely on the linear dust-\HI correlation to separate the dust from the contamination, and hence no masking is required to exclude pixels where dust and \NHI are poorly correlated. The algorithm recovers both \HI-correlated and uncorrelated dust emission by preserving the statistical properties of the contamination map, whereas the template-fitting approach recovers only the dust emission correlated with \HI.
\item We analyse the $\tr$ maps obtained from the three regions \Rlow, \Rint, and \Rhigh in terms of one-dimensional normalised histograms, angular power spectra, and three scalar MFs. All three $\tr$ maps are statistically consistent with one another and in close agreement with the $100\,\rsyn$ maps. No systematic correlation is found between $\tr$ and \NHI\ in either pixel or harmonic space, confirming that leakage of \HI-correlated dust emission into the $\tr$ map is negligible. The standard deviation of the $\tr$ map, computed over $32 \times 32$ pixel sub-regions, remains approximately constant over the range $0.8\times \hiunit < \NHI < 6.8\times \hiunit$, with $1\sigma$ fluctuations of $1.4\,\kJysr$.
\item Dust maps are used to analyze the relationship between dust and \HI emissions. The correlation coefficient between $\ts$ and the $\NLV$ map is $\geqslant 0.7$ across all three regions, confirming the strong dust-\HI correlation for local-velocity gas at high Galactic latitudes. In contrast, the correlation coefficient between $\ts$ and the $\NIV$ map is $\approx 0.2$ for regions \Rint and \Rhigh, and $-0.33$ for region \Rlow, indicating that dust associated with intermediate-velocity \HI is contributes weakly at high Galactic latitudes.
\item We characterize the spatial variations in dust emission relative to the hydrogen column density as a function of the angular scale. We perform the standard template-fit analyses on $\ts$ map using the two \HI templates, namely the \NLV and \NIV column-density maps over two different correlation scales. By varying $\Theta$ from $14.9\degree$ to $1.87\degree$, we progressively account for local dust emissivity variations within each sky region. We find that the mean value of the residual map ($\mathcal{E}$), after accounting for the local dust emissivity variations, approaches zero as $\Theta$ decreases and shows no dependence on \NHI, while its standard deviation within each sub-region scales approximately as $\langle \NHI \rangle^{0.7}$. These results indicate that the increasing variance of $\mathcal{E}$ with \NHI is consistent with spatial variations in the dust emissivity arising from combined variations in the dust temperature and spectral index, as well as the possible contribution of an additional ISM component not traced by the \HI\ templates, such as dust associated with $\molH$ and diffuse ionized gas.

\end{itemize}

This work opens up a promising avenue for extending the present study to the full sky at high Galactic latitudes. 
\begin{acknowledgments}
We thank Constant Auclair for the useful discussion and help during the start of the project. Some of the results in this paper have been derived using the \healpix package. We acknowledge use of the \Planck Legacy Archive. \texttt{HI4PI} is based on observations with the 100-m telescope of MPIfR at Effelsberg and the Parkes Radio Telescope, which is part of the Australia Telescope and is funded by the Common-wealth of Australia for operation as a National Facility managed by CSIRO. 
The computations in this paper were run on the GPU cluster at NISER supported by the Department of Atomic Energy of the Government of India. S.\ S.\ is supported by the National Postdoctoral Fellowship of the Science and Engineering Research Board (SERB), ANRF, Government of India (File No.: PDF/2023/000594).
\end{acknowledgments}
\software{
\healpix~\citep{Gorski:2005}, \texttt{reproject}~\citep{robitaille2020}, \texttt{PyWPH}~\citep{regaldo2021aa}, \texttt{PyTorch}~\citep{nocedal1980mc}, \namaster~\citep{alonso2019mnras}, \texttt{QuantImPy}~\citep{mantz2008jsm}
}
\bibliography{sed}{}
\bibliographystyle{aasjournalv7}
\appendix
\numberwithin{equation}{section}
\numberwithin{table}{section}
\numberwithin{figure}{section}
\section{Wavelet Phase Harmonics Coefficients}\label{app:wph-coeff}

We employ the WPH statistics originally introduced in~\cite{mallat2019jima} to characterize the complex structures present in two-dimensional non-Gaussian physical fields. These statistics have been previously applied to the analysis of various non-Gaussian physical fields~\citep{allys2020prd, zhang2021acha, jeffrey2021mnrasl, regaldo2023apj, auclair2024aa}. For completeness, we provide a brief description of the WPH \textit{auto}- and \textit{cross}-statistics used in this work. The WPH statistics are constructed from convolutions of the field $X$ with a set of pre-computed bandpass and low-pass filters (a wavelet transform), non-linear operations (phase harmonics transforms), and covariance estimates. The spatially localised complex-valued bump-steerable wavelets $\psi_{j,l}\paren*{\vec{r}}$, characterised by the dilation scale $j$ and rotation $l$, form the set of bandpass filters.

For two random fields $X$ and $Y$, the cross-WPH moments are
\begin{equation}
C^{\times}_{\lambda_{1},p_{1}, \lambda_{2}, p_{2}}\paren*{\vec{\tau}} = \Cov*{\left[X \ast \psi_{\lambda_{1}}\paren*{\vec{r}}\right]^{p_{1}}, \left[Y \ast \psi_{\lambda_{2}}\paren*{\vec{r}+\vec{\tau}}\right]^{p_{2}}},\label{eq:wph-cov}
\end{equation}
where $\lambda_{1}=\paren*{j_{1}, l_{1}}$ and $\lambda_{2}=\paren*{j_{2}, l_{2}}$ denote the oriented wavelet scales, and $\left[\,\cdot\,\right]^{p}$ is the pointwise $p$-th \textit{phase harmonic operator} on a complex field $z$, $z \mapsto \left[z\right]^{p} = \abs*{z}\,e^{i p \arg\paren*{z}}$. For stationary fields, and given a spatial translation vector $\vec{\tau}$, these moments quantify cross-scale and cross-orientation interactions between $X$ and $Y$ when their frequency supports overlap. The specific WPH statistics used are constructed from the $S^{p_{1}, p_{2}}_{\times}$ moments, which probe correlations within a single oriented scale ($\lambda_{1}=\lambda_{2}$), and the $C^{p_{1}, p_{2}}_{\times}$ moments, which characterize correlations between two oriented scales ($\lambda_{1}\neq\lambda_{2}$). These moments are
\begin{widetext}
\begin{eqnarray}
\begin{aligned}\label{eq:wph-coeff}
&S^{11}_{\times} = C^{\times}_{\lambda_{1},1,\lambda_{1},1}\paren*{\vec{\tau}} = \Cov*{X\ast\psi_{\lambda_{1}}\paren*{\vec{x}}, Y\ast\psi_{\lambda_{1}}\paren*{\vec{x} + \vec{\tau}}},\\
&S^{00}_{\times} = C^{\times}_{\lambda_{1},0,\lambda_{1},0}\paren*{\vec{\tau}}= \Cov*{\abs*{X\ast\psi_{\lambda_{1}}\paren*{\vec{x}}}, \abs*{Y\ast\psi_{\lambda_{1}}\paren*{\vec{x} + \vec{\tau}}}},\\
&S^{01}_{\times} = C^{\times}_{\lambda_{1},0,\lambda_{1},1}(\vec{\tau}) =  \Cov*{\abs*{X\ast\psi_{\lambda_{1}}\paren*{\vec{x}}}, Y\ast\psi_{\lambda_{1}}\paren*{\vec{x} + \vec{\tau}}},\\
&C^{00}_{\times} = C^{\times}_{\lambda_{1},0,\lambda_{2},0}(\vec{\tau}) = \Cov*{\abs*{X\ast\psi_{\lambda_{1}}\paren*{\vec{x}}}, \abs*{Y\ast\psi_{\lambda_{2}}\paren*{\vec{x} + \vec{\tau}}}},\\
&C^{01}_{\times} = C^{\times}_{\lambda_{1},0,\lambda_{2},1}(\vec{\tau}) = \Cov*{\abs*{X\ast\psi_{\lambda_{1}}\paren*{\vec{x}}}, Y\ast\psi_{\lambda_{2}}\paren*{\vec{x} + \vec{\tau}}},\\
&C^{\rm phase}_{\times} = C^{\times}_{\lambda_{1},1,\lambda_{2}, j_{1}/j_{2}}(\vec{\tau}) =  \Cov*{X\ast\psi_{\lambda_{1}}\paren*{\vec{x}}, \left[Y \ast \psi_{\lambda_{2}}\paren*{\vec{r}+\vec{\tau}}\right]^{j_{1}/j_{2}}}.
\end{aligned}
\end{eqnarray}
\end{widetext}
The corresponding auto-WPH moments are obtained by setting $X=Y$. In the remainder of this work, the superscript $\times$ is omitted when referring to auto-statistics.

We also use the \textit{scaling} moments $L^{\times}_{j, p_{1}, p_{2}}$ to probe the large-scale features of the fields that are not constrained by the WPH moments. The scaling moments are built using the Gaussian function  $\varphi_{j}\paren*{\vec{r}}$ as the low-pass filter:
\begin{equation}
L^{\times}_{j, p_{1}, p_{2}} = \Cov*{\left[X \ast \varphi_{j}\paren*{\vec{r}}\right]^{p_{1}}, \left[Y \ast \varphi_{j}\paren*{\vec{r}+\vec{\tau}}\right]^{p_{2}}}.
\end{equation}
We use $\paren*{p_{1}, p_{2}} : \lbrace \paren*{1,1}, \paren*{0,0}, \paren*{0,1} \rbrace $ for $L^{\times}_{j, p_{1}, p_{2}}$ 
and the auto-scaling moments are derived by choosing $X=Y$. We normalise the WPH and scaling moments relative to the reference fields $X_{0}$ and $Y_{0}$ as
\begin{eqnarray}
\begin{aligned}\label{eq:wph-coeff-norm}
    \tilde{C}_{\lambda_{1},p_{1},\lambda_{2},p_{2}}^\times(\vec{\tau}) &= \frac{\innerp*{X^{(\lambda_{1}, p_{1})}\paren*{\ve{r}}\overline{Y^{(\lambda_{2}, p_{2})}}\paren*{\ve{r}+\vec{\tau}}}}{\sqrt{\innerp*{\abs*{{X_{0}^{(\lambda_{1}, p_{1})}}}^{2}} \innerp*{\abs*{{Y_{0}^{(\lambda_{2}, p_{2})}}}^{2}}}},\\
    \tilde{L}_{j,p_{1}, p_{2}}^\times &= \frac{ \innerp*{ X^{(j,p_{1})}Y^{(j,p_{2})}}}{\sqrt{\innerp*{\abs*{ X_{0}^{(j,p_{1})}}^{2}}\innerp*{\abs*{Y_{0}^{(j,p_{2})}}^{2} }}},
\end{aligned}    
\end{eqnarray}
where $\innerp{\cdot}$ is the spatial average on $\vec{r}$, the overbar is the complex conjugate, and ${X^{(\lambda_{1},p_{1})} = \left[X\ast\psi_{\lambda_{1}}\right]^{p_{1}} - \innerp*{\left[X_{0}\ast\psi_{\lambda_{1}}\right]^{p_{1}}}}$. The normalisation depends on the reference maps $X_{0}$ and $Y_{0}$, and the choice of these reference maps has a significant impact on the component-separation algorithm. The normalised $S$- and $C$-moments are derived from $\tilde{C}^{\times}_{\lambda_{1},p_{1},\lambda_{2},p_{2}}$ by choosing the appropriate sets of $(\lambda_{1},p_{1},\lambda_{2},p_{2})$ specified in Eq.~(\ref{eq:wph-coeff}). The normalized scaling statistics are denoted by $\tilde{L}^{11}$, $\tilde{L}^{00}$, and $\tilde{L}^{01}$.

In this work, we consider fields of size $256 \times 256$ pixels and probe $J=6$ characteristic scales $2^{j}$ with $j\in \left[0,J-1\right]$. At each scale, we consider $L=4$ orientations between zero and $\pi$, defined by the angles $\frac{2\pi l}{L}$ relative to a reference direction. Here $\tau$ is the number of translation vector. With these parameters, we have 2468 WPH auto- and 9 scaling auto-statistics and 5096 WPH cross- and 12 scaling cross-statistics statistics. 

We use the publicly available Python package \texttt{PyWPH}\footnote{\href{https://github.com/bregaldo/pywph}{https://github.com/bregaldo/pywph}} to perform the GPU-accelerated WPH analysis~\citep{regaldo2021aa}.
\section{Validation on simulated Planck maps }\label{app:sim-results}

\begin{figure}[!htbp]
\centering
\includegraphics[width=\columnwidth]{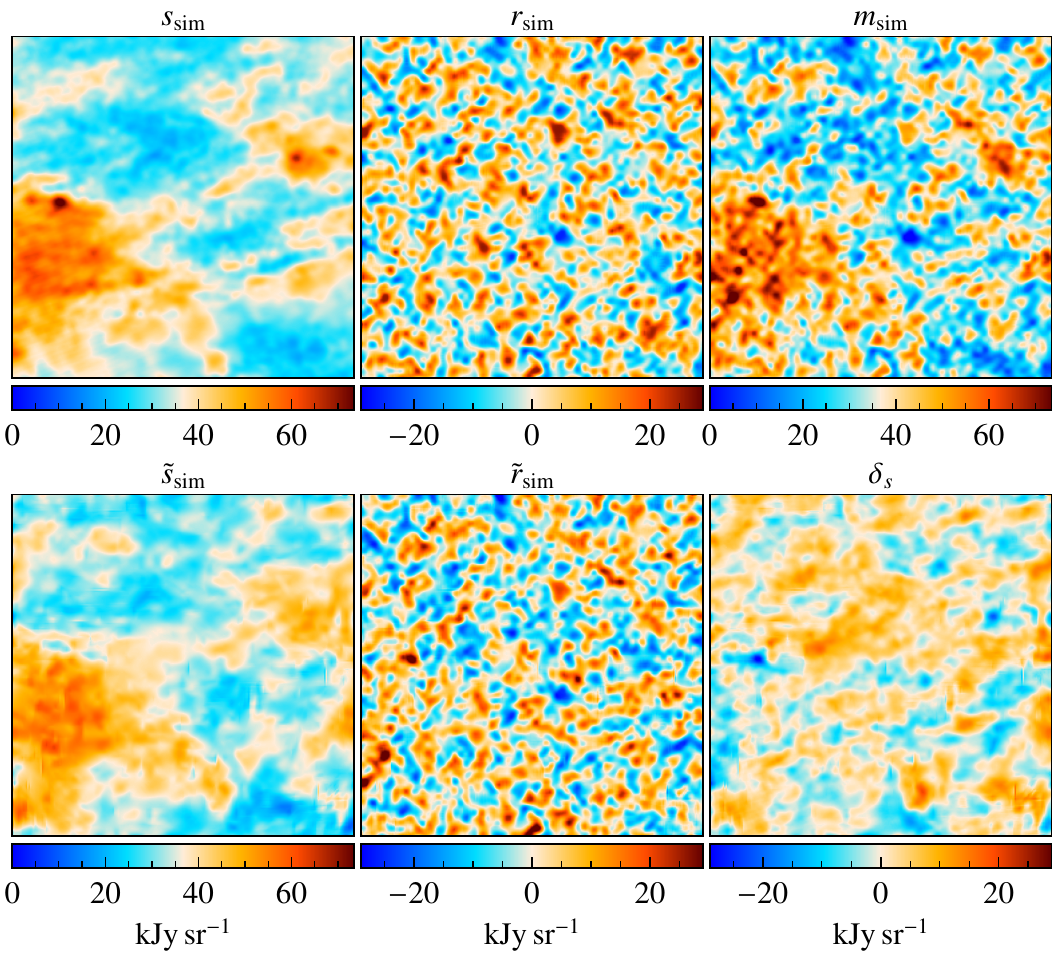}
\caption{\textit{Upper panel}: input maps at $\SNR=1$ at the \Rlow region. Columns from \textit{left to right} show the input dust map ($\ssim$), the input contamination map ($\rsim$) and the total simulated map ($\msim$). \textit{Lower panel}: Similarly, the component separated dust map ($\tssim$), the residual map ($\trsim$), and the difference between input and recovered dust map ($\delta_s$).}  
\label{fig:sim-wph-patch}
\end{figure}
We validate our component-separation algorithm using mock \Planck data for the \Rlow region at $\SNR=1$. Following~\citetalias{sinha2026aa}, we use the scaled $\scfd$ map as the input dust signal ($\ssim$) and add a random realisation of contamination ($\rsim$), drawn from the $100\,\rsyn$ realisations to construct the mock data map ($\msim$). The resulting $\msim$ map, dominated by small-scale fluctuations, closely resembles the \Planck data $m$ map shown in Fig.~\ref{fig:data-result-map-R1}. For the simulated map in this region, the correlation coefficient $\rho$ between $\msim$ and $\NHI$ is $0.73$, and that between $\ssim$ and $\NHI$ is $0.90$. We then apply the component-separation algorithm described in Sect.~\ref{sec:wph-algorithm} to recover the dust and contamination maps. In practice, the true dust signal is unknown and therefore cannot be used to normalise the WPH coefficients. Instead, the coefficients must be normalized using a reference map that provides a close approximation to the true dust emission. Since the $\scfd$ map differs slightly from the observed \Planck dust map in the \Rlow region (see Sect.~\ref{sec:data-results}), we normalize the WPH coefficients using the same $\sref$ map as in Sect.~\ref{sec:wph-algorithm}. This simulation therefore allows us to quantify the bias introduced by using an approximate reference map in place of the true dust map for the normalization of the WPH coefficients.

\begin{figure}[!htbp]
\centering
\includegraphics[width=\columnwidth]{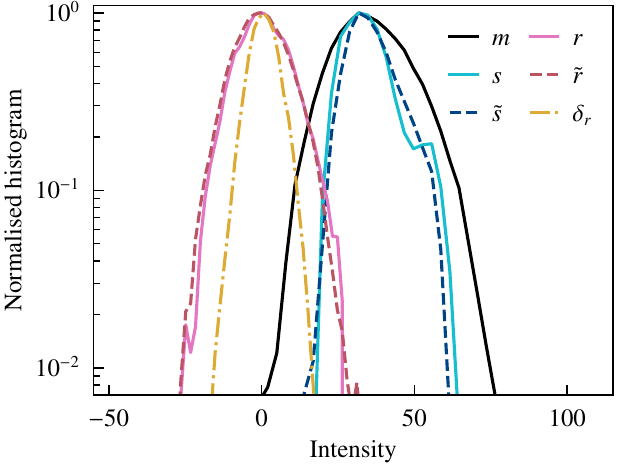}
\caption{One dimensional normalised distribution of $m$ (black solid line), $\ssim$ (cyan solid line), $\tssim$ (blue dashed line), $\rsim$ (pink solid line) and $\trsim$ (red dashed line). The plot shows that the recovered dust and contamination components successfully reproduce the distributions of their corresponding input components. The distribution of $\delta_s$ (orange dot-dashed line) is distinct from that of $\rsim$ and is instead consistent with a random statistical fluctuation.}
\label{fig:sim-wph-hist} 
\end{figure}

Figure \ref{fig:sim-wph-patch} shows the $\ssim$, $\rsim$ and $\msim$ in the upper panel, in $\kJysr$. The lower panel shows the recovered dust map after the component-separation $\tssim$, the recovered contamination map $\trsim$ and the difference between the recovered and the input dust maps ($\delta_{s} =\tssim-\ssim$). After component separation, the correlation coefficient $\rho$ between $\tssim$ and $\NHI$ becomes 0.90, which is close to the input value. The recovered value of $\rho$ highlights the fact the most of the small-scale power of the dust emission are reconstructed well that are buried under the contamination.  Figure~\ref{fig:sim-wph-hist} compares the normalised histograms of the input and recovered components. The black solid line represents the distribution of $\msim$, which is significantly broader than that of $\ssim$ (cyan solid line) because the emission in this region is contamination dominated. After the component-separation, the non-Gaussian distribution of the dust signal is successfully recovered by $\tssim$ (blue dashed line). As expected, the distributions of $\rsim$ (pink solid line) and $\trsim$ (red dashed line) are both close to Gaussian and centred around zero. The narrow distribution of $\delta_s$ (orange dot-dashed line) is also symmetric about zero, indicating that the residuals are random and do not resemble the contamination. 

To further test whether the recovered $\trsim$ map deviates from the expected statistical properties of the contamination, we also compute the MFs of the $\trsim$ map and and compare them with the corresponding distributions obtained from the $100\,\rsyn$ realisations. We find that the MFs of $\trsim$ are statistically consistent with those of the $\rsyn$ maps and show no significant deviations. Since these results are fully consistent with the expected Gaussian distribution of the $\rsyn$ maps, we do not show the corresponding plots for brevity.
\begin{figure*}[!htbp]
\centering
\includegraphics[width=0.8\linewidth]{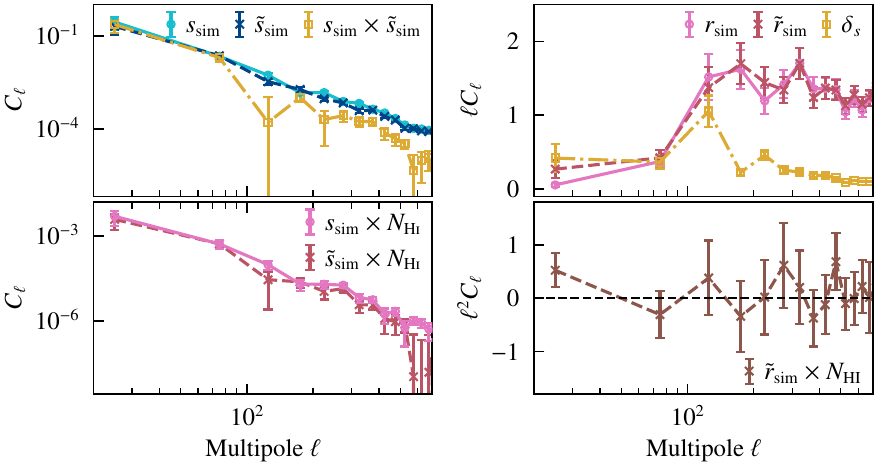}
\caption{The first column shows the power spectra comparison of the input and the recovered dust maps. The \textit{upper panel} shows $\Cl$  for the $\ssim$ (cyan solid line), $\tssim$ (blue dashed line) and the cross spectrum of $\ssim \times \tssim$ (orange dot-dashed line). The \textit{lower panel} shows the cross spectra $\ssim$ (pink solid line) and $\tssim$ (red dashed line) with the \NHI map. The second column shows the power spectra comparison of the input and the recovered contamination maps. The \textit{upper panel} shows $\ell\Cl$ with $\ell$ for the $\rsim$ (pink solid line), $\trsim$ (red dashed line) and $\delta_s$ (orange dot-dashed line). The \textit{lower panel} shows the cross spectra in $\ell^{2}\Cl$ with $\ell$ of $\trsim$ (brown dashed line) with the \NHI map.}
\label{fig:sim-wph-ps} 
\end{figure*}
\begin{figure*}[!htbp]
\centering
\includegraphics[width=0.9\linewidth]{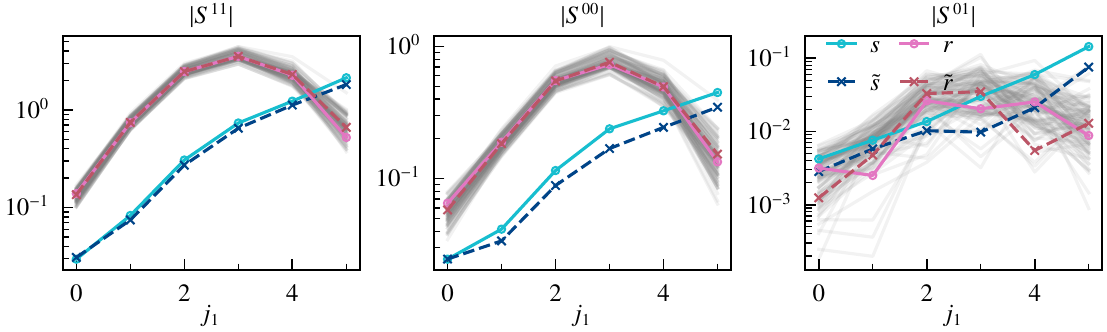}
\caption{Absolute values of the WPH coefficients $S^{11}$, $S^{00}$, and $S^{01}$ for the input and recovered maps. The input and recovered dust maps, $\ssim$ and $\tssim$, are shown by cyan circles with a solid line and blue crosses with a dashed line, respectively. The corresponding input and recovered contamination maps are showed by magenta circles ($\rsim$) and red crosses ($\trsim$), respectively. The gray lines correspond to the coefficients obtained from the $100\,\rsyn$ maps.} 
\label{fig:sim-wph-coeff} 
\end{figure*}

Next, we compute the angular power spectra of the input and recovered maps. The first column of Fig.~\ref{fig:sim-wph-ps} compares $\Cl$ of $\ssim$ (cyan solid line), $\tssim$ (blue dashed line), and their cross-power spectrum (orange dot-dashed line) in the upper panel. The $\tssim$ spectrum closely matches that of the $\ssim$ over the full multipole range, while the cross-power spectrum shows a slight decorrelation at the smallest angular scales. 

The lower left panel of Fig.~\ref{fig:sim-wph-ps} shows the cross-power spectra of the $\ssim$ and $\tssim$ maps with the \NHI map. Since the algorithm exploits the dust-\NHI correlation, the recovered dust map is expected to preserve this cross-spectrum. 

The second column of Fig.~\ref{fig:sim-wph-ps} shows the power spectra of $\rsim$, $\trsim$, and $\delta_s$ in the upper panel. The recovered contamination spectrum agrees well with the input spectrum over all angular scales. In contrast, the power spectrum of $\delta_s$ remains significantly below those of $\rsim$ and $\trsim$ across almost the entire multipole range, indicating that $\delta_s$ represents a statistical difference between $\ssim$ and $\tssim$ rather than residual contamination.

The lower right panel shows the cross-power spectrum of $\trsim \times \NHI$, plotted in $\ell^2\Cl$, which does not exhibit a significant  statistical correlation, confirming that the WPH-based component-separation method successfully recovers the dust and contamination components while suppressing leakage of dust emission into the recovered contamination map.

Figure~\ref{fig:sim-wph-coeff} presents the absolute values of the WPH coefficients ($S^{11}$, $S^{00}$, and $S^{01}$), computed with $J=6$ and averaged over the $L$ orientations, for the input and recovered maps. The gray lines correspond to the coefficients measured from the $100\,\rsyn$ maps. The $S^{11}$ coefficients of the input and recovered dust maps, $\ssim$ (blue circles) and $\tssim$ (cyan crosses), are in good agreement, while the $\tssim$ map exhibits slightly less large-scale power in $S^{00}$ than the input $\ssim$ map. Their $S^{01}$ coefficients show small but noticeable differences, indicating that part of the phase-coupling information is not fully recovered. For the contamination component, the coefficients of both $\tr$ (pink circles) and $\trsim$ (red crosses) agree well in $S^{11}$ and $S^{00}$. Although differences are also present in their $S^{01}$ coefficients, they remain well within the distribution of coefficients obtained from the $100\,\rsyn$ realisations.
\section{Comparison of different algorithms}\label{app:data-method-comp}

\begin{figure}[!htbp]
\centering
\includegraphics[width=0.8\columnwidth]{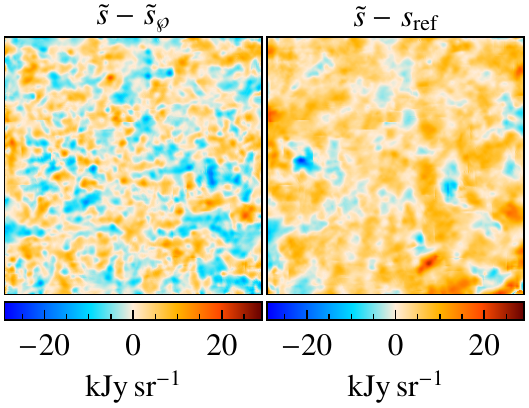}
\caption{The differences between the dust maps recovered by the three component-separation methods for the \Rlow region.}  
\label{fig:data-map-comp}
\end{figure}

Figure~\ref{fig:data-map-comp} shows the differences between the dust maps recovered by the three component-separation methods for the common \Rlow region: $\ts-\tssc$ (left column) and $\ts-\sref$ (right column). The difference map involving the $\tssc$ map, derived using the methodology of \citetalias{sinha2026aa}, clearly reveals residual small-scale fluctuations, whereas the difference with the template-fit map, $\sref$, is comparatively smooth.

\begin{figure}[!htbp]
\centering
\includegraphics[width=\columnwidth, keepaspectratio=True]{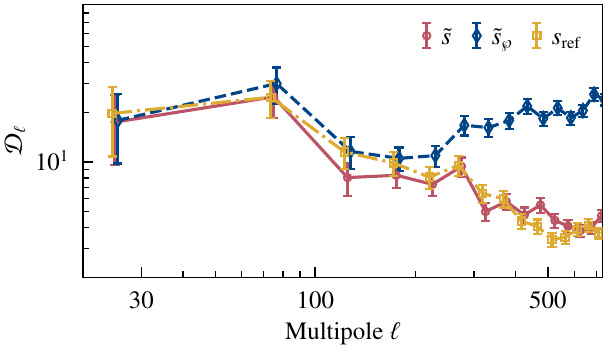}
\caption{Recovered dust power spectra in $\mathcal{D}_{\ell}$ computed from region $\Rlow$ using three component-separation methods: this work (red circles with solid line),~\citetalias{sinha2026aa} (blue diamonds with dashed line), and the template-fit approach (orange squares with dot-dashed line). The power spectrum of $\tssc$ shows clear deviation from those of $\ts$ and $\sref$ for $\ell > 175$. The data points are shifted slightly along the $x$-axis.}
\label{fig:data-result-cl-comp}
\end{figure}

Figure~\ref{fig:data-result-cl-comp} presents the angular power spectra of the three recovered dust maps, expressed as $\Dl = \ell(\ell+1)\Cl/(2\pi)$. The power spectrum of $\ts$ agrees well with $\sref$ up to $\ell = 700$ within $1\sigma$ uncertainties, confirming that the dust signal extracted in this work is consistent with the template-fit approach. In contrast, the $\tssc$ map shows spurious excess power at small angular scales above $\ell = 175$ compared to the other two methods. At large angular scales ($\ell < 175$), all three dust power spectra agree well with each other. 

\begin{figure}[!htbp]
\centering
\includegraphics[width=\columnwidth, keepaspectratio=True]{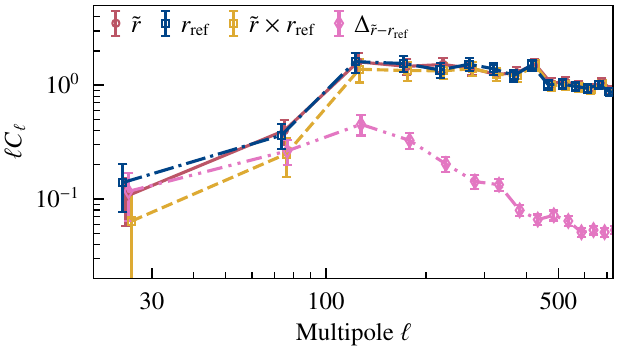}
\caption{Shows the auto-power spectra in $\ell\Cl$ of $\tr$ (red circles) and $\rref$ (blue squares), along with their cross-power spectrum (orange squares). The close agreement between the three spectra indicates that the two maps trace the same underlying structures. The power spectrum of the $\Delta_{\tr-\rref}$ difference map (magenta diamonds) retains only large-scale power and decreases rapidly towards smaller scales.}\label{fig:data-cib-ps-hmc}
\end{figure}

Figure~\ref{fig:data-cib-ps-hmc} shows the auto-power spectra of $\tr$ and $\rref$, along with their cross-power spectrum, $\tr \times \rref$. The close agreement between the auto- and cross-spectra, particularly at small angular scales, demonstrates that the contamination maps recovered by the two component-separation methods trace the same underlying structures and retain the same phase information. This indicates that $\tr$ is not merely a statistical realisation of the contamination but a physically meaningful reconstruction. The small deviations between the auto- and cross-spectra are confined primarily to large angular scales. This is also evident in the power spectrum of the corresponding difference map ($\Delta_{\tr-\rref} \equiv \tr-\rref$), which retains power on large scales ($\ell < 100$) but decreases rapidly towards smaller scales, indicating that the discrepancies between the two reconstructions are dominated by large-scale modes. On these large angular scales, the template-fitting approach, which assumes a constant emissivity within $1.8\degree$ patches, exhibits phase differences relative to $\tr$, as reflected in their cross-power spectrum. Since the WPH-based component-separation method is not restricted by a fixed correlation scale, it is able to recover large-scale structures that are not captured by the template-fitting reconstruction $\rref$.

\begin{figure}[!htbp]
\centering
\includegraphics[width=\columnwidth, keepaspectratio=True]{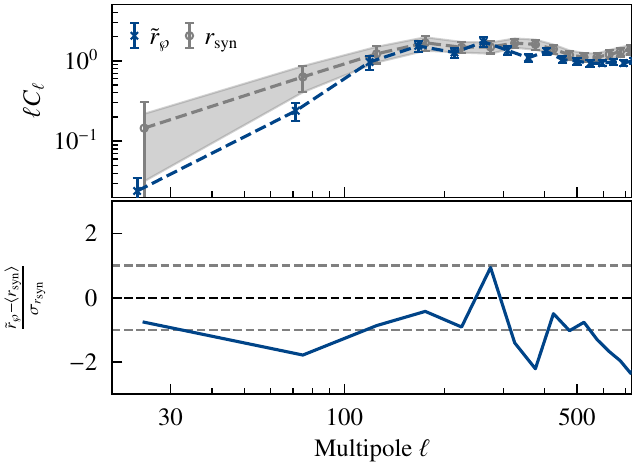}
\caption{\textit{Upper panel}: shows the power spectra in $\ell\Cl$ of $\trsc$ (blue crosses), together with the mean (gray line) and $1\sigma$ uncertainty band of the $100\,\rsyn$ realisations. The spectrum obtained using the methodology of~\citetalias{sinha2026aa} deviates significantly from the $1\sigma$ uncertainty band at all scales. The blue crosses are shifted slightly along the $x$-axis. \textit{Lower panel}: shows the deviation of $\trsc$ from the mean $\rsyn$ spectrum relative to the $1\sigma$ uncertainty.}
\label{fig:data-cib-ps-s26}
\end{figure}

The upper panel of Fig.~\ref{fig:data-cib-ps-s26} shows the power spectrum in $\ell\Cl$ of $\trsc$, obtained using the methodology of~\citetalias{sinha2026aa}, together with those of the $100\,\rsyn$ realisations. The gray band represents the $1\sigma$ uncertainty derived from the $100\,\rsyn$ realisations. The power spectrum of $\trsc$ deviates significantly from the expected range at small angular scales, with most of the small-scale power lying outside the $1\sigma$ uncertainty band, indicating that these scales are not accurately recovered. The lower panel shows the deviation of $\trsc$ from the mean $\rsyn$ spectrum outside the $1\sigma$ uncertainty range.
\end{document}